\documentclass[twocolumn,aps,floatfix]{revtex4-1}
\usepackage{amsmath,amssymb}
\usepackage{graphicx} 
\usepackage{dcolumn} 
\usepackage{bm} 
\usepackage{braket} 
\usepackage[usenames,dvipsnames]{color}
\usepackage[
colorlinks=true, citecolor=blue, urlcolor=blue, linkcolor=blue,
setpagesize=false, bookmarks=false
]{hyperref}

\begin{document}


\title{Photoinduced electron-electron pairing in the extended Falicov-Kimball model}
\author{Ryo Fujiuchi$^{1}$}
\author{Tatsuya Kaneko$^{2}$}
\author{Yukinori Ohta$^{1}$}
\author{Seiji Yunoki$^{2,3,4}$}
\affiliation{
$^1$Department of Physics, Chiba University, Chiba 263-8522, Japan\\
$^2$Computational Condensed Matter Physics Laboratory, RIKEN Cluster for Pioneering Research (CPR), Wako, Saitama 351-0198, Japan\\
$^3$Computational Quantum Matter Research Team, RIKEN Center for Emergent Matter Science (CEMS), Wako, Saitama 351-0198, Japan\\
$^4$Computational Materials Science Research Team, RIKEN Center for Computational Science (R-CCS), Kobe, Hyogo 650-0047, Japan 
}
\date{\today}


\begin{abstract}  
By employing the time-dependent exact diagonalization method,  
we investigate the photoexcited states of the excitonic insulator in the extended Falicov-Kimball model (EFKM).
We here show that the pulse irradiation can induce the interband electron-electron pair correlation in the photoexcited 
states, while the excitonic electron-hole pair correlation in the initial ground state is strongly suppressed.
We also show that the photoexcited states contains the eigenstates of the EFKM with 
a finite number of interband electron-electron pairs, which are responsible for the enhancement of the electron-electron 
pair correlation. 
The mechanism found here is due to the presence of the internal SU(2) pairing structure in the EFKM and thus it is essentially 
the same as that for the photoinduced $\eta$-pairing in the repulsive 
Hubbard model reported recently [T. Kaneko {\it et al.}, Phys. Rev. Lett. {\bf 122}, 077002 (2019)]. 
This also explains why the nonlinear optical response is effective to induce the electron-electron pairs in the 
photoexcited states of the EFKM. 
Furthermore, we show that, unlike the $\eta$-pairing in the Hubbard model, the internal SU(2) structure is 
preserved even for a nonbipartite lattice when the EFKM has the direct-type band structure, in which 
the pulse irradiation can induce the electron-electron pair correlation with momentum $\bm{q}=\bm{0}$ in the 
photoexcited states. 
We also discuss briefly the effect of a perturbation that breaks the internal SU(2) structure.  
\end{abstract}

\maketitle



\section{Introduction}

Physics of the excitonic order and excitonic insulator~\cite{JRK67,Ko67,HR68} has attracted renewed 
attention~\cite{Ku15,NWNetal16,KO16}, triggered by recent discoveries of a number of candidate materials.
The excitonic order is described as a quantum condensed state of electron-hole pairs (or excitons) via interband 
Coulomb interactions~\cite{JRK67,Ko67,HR68}, and the insulator realized by the excitonic order or strong excitonic 
correlation is called the excitonic insulator.
As the promising candidates among transition-metal compounds, the possible realization of spin-singlet excitonic 
phase has been suggested in the transition-metal chalcogenides 
1$T$-TiSe$_2$~\cite{CMCetal07,MCCetal09,MBCetal11,ZFBetal13,WSY15,KOY18} and 
Ta$_2$NiSe$_5$~\cite{WSTetal09,KTKetal13,*KTKetal13e,SWKetal14,SKO16,LKLetal17,SNKetal18}.

Recently, the pump-probe measurements are applied to these candidate materials~\cite{RHWetal11,MJBetal11,HRKetal12,PLMetal14,MEUetal16,MPNetal16,MHGetal17,MHNetal18,WTHetal18,WTAetal18,OOSetal18}, 
and the nonequilibrium dynamics of the excitonic insulators induced by laser pulse have also been investigated 
theoretically~\cite{GWE16,MGEetal17,TDY18,TSO18}. 
In 1$T$-TiSe$_2$,  the pump-probe measurements have been used to extract the excitonic contribution from the electron-phonon coupled charge density wave state~\cite{RHWetal11,MJBetal11,HRKetal12,PLMetal14,MEUetal16,MPNetal16}. 
In Ta$_2$NiSe$_5$, the pump fluence dependent gap narrowing and opening~\cite{MHGetal17}, coherent order parameter oscillations~\cite{WTHetal18,WTAetal18}, and insulator-to-metal transition~\cite{OOSetal18} have been observed as 
indications of an excitonic order. 
Concurrently with the experiments, the theories for the photoinduced dynamics of the excitonic insulator 
have been developed by using the Hartree-Fock and GW approximations~\cite{GWE16,MGEetal17,TDY18,TSO18}.
However, since these theoretical studies employed the approximations, the numerically exact analysis 
based on unbiased methods is desirable in order to provide new insight for the photoinduced dynamics 
of the excitonic insulator. 

Here, in this paper, we employ the time-dependent exact diagonalization method to 
investigate the pulse excited states of the extended Falicov-Kimball model (EFKM), which is the simplest spinless 
model for describing the excitonic 
insulator~\cite{Ba02,*Ba03e,IPBetal08,SEO11,ZIBetal12,KEFetal13,EKOetal14,HKMetal17}.
In particular, we demonstrate that the interband electron-electron pair correlation can be photoinduced 
in the excitonic insulator 
of the EFKM, in analogy with the photoinduced $\eta$-pairing in the Hubbard model, 
where the pair density wave like correlation is induced by the pulse irradiation in the Mott insulator~\cite{KSSetal19}. 
By decomposing the photoexcited states into the eigenstates of the 
EFKM, we show that the photoexcited states have a finite weight of the eigenstates with a finite number of 
electron-electron pairs, 
thus enhancing the electron-electron pair correlation in the photoexcited states. 
The mechanism found here is due to the presence of the internal SU(2) pairing structure in the EFKM, which is 
in principle the same as that for the photoinduced $\eta$-pairing in the Hubbard model~\cite{KSSetal19}. 
Furthermore, we show that, in contrast to the $\eta$-pairing in the Hubbard model, this internal SU(2) structure 
is preserved even for a nonbipartite lattice when the EFKM has the direct-type electron and hole band structure, 
in which the electron-electron pair correlation with momentum $\bm{q}=\bm{0}$ can be induced by 
the pulse irradiation. 
 
The rest of this paper is organized as follows.  
In Sec.~\ref{sec:model}, we introduce the EFKM, and discuss the internal SU(2) structure of the model 
and the relation to the Hubbard models. 
In Sec.~\ref{sec:method}, we briefly describe the numerical method to calculate the dynamics of 
the time-dependent Hamiltonian. 
In Sec.~\ref{sec:results}, we provide the numerical results for the one-dimensional (1D) chain and the 
two-dimensional (2D) square and triangular lattices.   
The paper is concluded in Sec.~\ref{sec:summary}. 
The photoinduced interband $\eta$-pairing is discussed for the EFKM with the indirect-gap-type band structure in 
Appendix~\ref{sec:app-eta}.


\section{Model}\label{sec:model}

\subsection{Extended Falicov-Kimball model (EFKM)}

To study the effects of photoexcitation in an excitonic insulator, we consider the EFKM at half filling.  
The model is defined by the following Hamiltonian: 
\begin{align}
{\hat {\mathcal{H}}}  =  &- \sum_{\langle i, j \rangle} \sum_{\alpha=1,2} t^{(\alpha)}_h  \left( {\hat c}_{i,\alpha}^{\dag}{\hat c}_{j,\alpha} + {\rm H.c.} \right)   
\notag \\
&+ \frac{D}{2} \sum_{j=1}^L  \left( {\hat n}_{j, 2} -{\hat n}_{j, 1} \right)  + U \sum_{j=1}^L  {\hat n}_{j, 1}  {\hat n}_{j, 2}    ,
\label{PI-EFKM_eq_H}                    
\end{align}
where ${\hat c}_{j,\alpha}$ (${\hat c}_{j,\alpha}^{\dag}$) is the annihilation (creation) operator of 
an electron at site $j$ with orbital $\alpha$~($= 1, 2$), and 
${\hat n}_{j, \alpha}={\hat c}_{j,\alpha}^{\dag}{\hat c}_{j,\alpha}$. 
The sum indicated by $\langle i, j \rangle$ runs over all pairs of nearest-neighbor sites $i$ and $j$ 
with the hopping parameter $t^{(\alpha)}_h$ that depends on the orbital. 
$D$ ($>0$) is the energy level splitting between the two orbitals and $U$ ($>0$) is the interband repulsive 
interaction, which gives rise to the strong electron-hole pair (i.e., exciton) correlation. 
$L$ is the number of lattice sites, and 
$N_{\alpha}$ is the total number of electrons for each orbital $\alpha\,(=1,2)$. 

\begin{figure}[t]
\begin{center}
  \includegraphics[width=0.9\columnwidth]{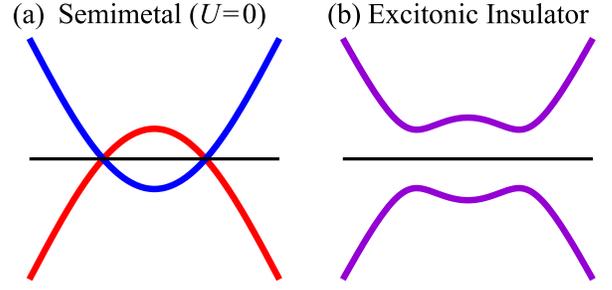}
\caption{
Schematic band structures of (a) a semimetal ($U=0$) and (b) an excitonic insulator 
in the EFKM with $t_h^{(1)} \cdot t_h^{(2)} < 0$.  
}
\label{fig:efkm_sche}
\end{center}
\end{figure}

The sum of the first and second terms of Eq.~(\ref{PI-EFKM_eq_H}) may be written in momentum 
($\bm{k}$) space as 
\begin{align}
\hat {\mathcal{H}}_0=\sum_{\bm{k},\alpha} \epsilon_{\alpha}(\bm{k}){\hat c}_{\bm{k},\alpha}^{\dag}{\hat c}_{\bm{k},\alpha}
\end{align}
with 
\begin{align}
\epsilon_{1}(\bm{k})  = -2t^{(1)}_h \sum_{\tau} \cos k_{\tau} - \frac{D}{2}
\end{align}
and
\begin{align}
\epsilon_{2}(\bm{k})  = -2t^{(2)}_h \sum_{\tau} \cos k_{\tau} + \frac{D}{2}, 
\end{align}
where $k_\tau = \bm{k}\cdot \bm{a}_{\tau}$ and $\bm{a}_\tau$ is the vector between the nearest-neighbor sites 
$i$ and $j$.  
Here, we implicitly assume that the hoppings are finite between sites connected through the primitive translation 
vectors and the unit cell contains only a single site. 
Figure~\ref{fig:efkm_sche}(a) shows a schematic band structure of the EFKM with $t_h^{(1)} \cdot t_h^{(2)} < 0$ 
and $U=0$, which is a direct-gap-type semimetal~\cite{direct}.
At half filling, i.e., $N_1 + N_2=L$, the ground state of the EFKM for large $U$ is an insulator 
[see Fig.~\ref{fig:efkm_sche}(b)] with 
the strong excitonic correlation~\cite{SNKetal18}. 
Note that, when $t_h^{(1)} = t_h^{(2)}$, the EFKM is essentially equivalent to the Hubbard model. 
Therefore, as in the case of the Hubbard model, the EFKM with $t_h^{(1)} = t_h^{(2)}$ has the internal SU(2) structure 
defined by the $\eta$-pairing operators~\cite{Ya89,EKS91,EFGetal05}. 
Below, we will show that the EFKM with $t_h^{(1)} = - t_h^{(2)}$ 
displays the different internal SU(2) structure defined by interband electron-electron pairing operators, 
which we refer to as $\Delta$-pairing operators.  Most importantly, this internal SU(2) structure is realized  even for 
nonbipartite lattices and therefore it is not simply obtained by a local gauge transformation form the 
Hubbard model (see Sec.~\ref{sec:rhm}). 

\subsection{Internal SU(2) structure in EFKM}

In order to consider the interband electron-electron pairing in the EFKM, 
let us first introduce the following operators: 
\begin{align}
\hat{\Delta}^+_j= \hat{c}^{\dag}_{j,2} \hat{c}^{\dag}_{j,1}, 
\;\; 
\hat{\Delta}^-_j= \hat{c}_{j,1} \hat{c}_{j,2}, 
\end{align}
and 
\begin{align}
\hat{\Delta}^z_j=\frac{1}{2}\left( {\hat n}_{j, 1} + {\hat n}_{j, 2} -1 \right). 
\end{align}
We can easily show that these operators satisfy the SU(2) commutation relations, i.e., 
\begin{align}
&\left[ \hat{\Delta}^+_j, \hat{\Delta}^-_j \right] = 2 \hat{\Delta}^z_j , 
\\
&\left[ \hat{\Delta}^z_j,  \hat{\Delta}^\pm_j \right] = \pm \hat{\Delta}^\pm_j .
\end{align}
Similarly, we introduce the total $\hat{\Delta}$ operators as 
\begin{align}
\hat{\Delta}^+= \sum_{j} \hat{c}^{\dag}_{j,2} \hat{c}^{\dag}_{j,1} = \sum_{\bm{k}} \hat{c}^{\dag}_{-\bm{k},2} \hat{c}^{\dag}_{\bm{k},1}, 
\\
\hat{\Delta}^-= \sum_{j} \hat{c}_{j,1} \hat{c}_{j,2} = \sum_{\bm{k}} \hat{c}_{\bm{k},1} \hat{c}_{-\bm{k},2}, 
\end{align}
and 
\begin{align}
\hat{\Delta}_z=\frac{1}{2} \sum_{j} \left( {\hat n}_{j, 1} + {\hat n}_{j, 2} -1 \right),  
\label{eq:delta_z}
\end{align}
which also satisfy the SU(2) commutation relations, i.e.,  
\begin{equation}
\left[ \hat{\Delta}^+, \hat{\Delta}^- \right] = 2 \hat{\Delta}^z, \quad
\left[ \hat{\Delta}^z,  \hat{\Delta}^\pm \right] = \pm \hat{\Delta}^\pm ,
\label{eq:su(2)}
\end{equation}
and are referred to as $\Delta$-pairing operators. 
Defining the total $\Delta$-pairing operator as 
\begin{align}
{\hat{\Delta}}^2=\frac{1}{2} ( \hat{\Delta}^+\hat{\Delta}^- + \hat{\Delta}^-\hat{\Delta}^+  ) + \hat{\Delta}_z^2,
\end{align} 
we can also easily show that 
\begin{align}
\left[ {\hat{\Delta}}^2,  \hat{\Delta}_z \right] = 0
\label{eq:cr_delta}
\end{align}
 
The essential property of the $\Delta$-pairing operators is 
\begin{align}
\left[ \hat{\mathcal{H}}_0,\hat{\Delta}^{+} \right]  
&=  \sum_{\bm{k}} \left[ \epsilon_{1}(\bm{k})   + \epsilon_{2}(-\bm{k})  \right] \hat{c}^{\dag}_{-\bm{k},2} \hat{c}^{\dag}_{\bm{k},1}
\notag \\
& = - 2 \left( t^{(1)}_h +t^{(2)}_h \right)  \sum_{\tau,\bm{k}} \cos(k_{\tau}) \, \hat{c}^{\dag}_{-\bm{k},2} \hat{c}^{\dag}_{\bm{k},1}  
\label{PI-EFKM_eq_H-Delta} 
\end{align}
and therefore $[\hat{\mathcal{H}}_0,\hat{\Delta}^{+}] =0$ when $ t^{(1)}_h = -t^{(2)}_h$.  
A similar relation holds for $\hat{\Delta}^{-}$ and thus  
$[\hat{\mathcal{H}}_0,\hat{\Delta}^{\pm}] =0$ when $ t^{(1)}_h = -t^{(2)}_h$.  
The commutation relation for the third term of Eq~(\ref{PI-EFKM_eq_H}), 
$\hat{\mathcal{H}}_U=U \sum_{j}  {\hat n}_{j, 1}  {\hat n}_{j, 2}$, is given by 
$[\hat{\mathcal{H}}_U,\hat{\Delta}^{\pm}] = \pm U \hat{\Delta}^{\pm}$.  Hence, we have the relation 
\begin{align}
\left[ \hat{\mathcal{H}},\hat{\Delta}^{\pm} \right] = \pm U \hat{\Delta}^{\pm} 
\label{PI-EFKM_eq_HEFKM-Delta}   
\end{align}
for the EFKM with $t^{(1)}_h = -t^{(2)}_h$.  
Using this commutation relation and the definition of $\hat{\Delta}_z$ in Eq.~(\ref{eq:delta_z}), we can show that 
$\hat{\mathcal{H}}$ commutes with $\hat{\Delta}^2$ and $\hat{\Delta}_z$, i.e., 
\begin{align}
\left[ \hat{\mathcal{H}}, \hat{\Delta}^2 \right] = \left[ \hat{\mathcal{H}}, \hat{\Delta}_z \right]=0, 
\label{eq:cr_h}
\end{align}
when $t^{(1)}_h = -t^{(2)}_h$.  
In this paper,  
we refer to a model as preserving the internal SU(2) structure with respect to the $\Delta$-pairing 
operators, if the model described by Hamiltonian $\hat{\mathcal{H}}$ satisfies the commutation relations 
given in Eq.~(\ref{eq:cr_h}) with the $\Delta$-pairing operators that themselves satisfy the SU(2) commutation 
relations in Eq.~(\ref{eq:su(2)})~\cite{SU2}.

Eqs.~(\ref{eq:cr_delta}) and (\ref{eq:cr_h}) imply that any eigenstate of $\hat{\cal{H}}$ is also the eigenstate of 
${\hat{\Delta}}^2$ and $\hat{\Delta}_z$ 
with eigenvalues $\Delta(\Delta+1)$ and $\Delta_z$, respectively~\cite{delta}.  
We denote this eigenstate as $|\Delta, \Delta_z\rangle$.  
Assuming that $N_1 \ge N_2$ and 
$L-N_1+N_2$ is even, 
$|\Delta, \Delta_z\rangle $ can take 
$\Delta=0,1,2,\cdots , (L-N_1+N_2)/2$
and $\Delta_z=-\Delta,-\Delta+1, \cdots, \Delta-1, \Delta$. 
Note that $\Delta_z=0$ at half filling with $N_1 + N_2 = L$. 
The state $|\Delta, \Delta_z = - \Delta \rangle $ is the lowest weight state (LWS) that 
satisfies $\hat{\Delta}^{-}|\Delta, \Delta_z = - \Delta \rangle =0$~\cite{EKS91,EFGetal05}.
The other eigenstates with $\Delta$ can be generated from the LWS by applying $\hat{\Delta}^{+}$. 
For example, the eigenstate with finite $\Delta$ ($>0$) at half filling $\Delta_z=0$ is given as 
$|\Delta, \Delta_z =  0 \rangle \propto ( \hat{\Delta}^{+} )^{\Delta} |\Delta, \Delta_z =  -\Delta \rangle$, 
indicating that a $\Delta$-pairing state is generated from a hole-doped state (i.e, $\Delta_z < 0$). 
Note also that, because of Eq.~(\ref{PI-EFKM_eq_HEFKM-Delta}), the energy is increased (decreased) by $U$ 
every time that $\hat{\Delta}^+$ ($\hat{\Delta}^-$) is applied to the eigenstate of $\hat{\mathcal{H}}$.

Similarly, the EFKM with $t^{(1)}_h = t^{(2)}_h$ (i.e., the indirect-gap-type band structure) has 
the internal SU(2) structure with respect to the interband $\eta$-pairing operators defined as 
$\hat{\eta}^+= \sum_{j} (-1)^j \hat{c}^{\dag}_{j,2} \hat{c}^{\dag}_{j,1}$, 
$\hat{\eta}^-= \sum_{j} (-1)^j \hat{c}_{j,1} \hat{c}_{j,2}$, and 
$\hat{\eta}_z=\frac{1}{2} \sum_{j} \left( {\hat n}_{j, 1} + {\hat n}_{j, 2} -1 \right)$. 
The details are discussed in Appendix~\ref{sec:app-eta}.

\subsection{External field}

The time-dependent external field is introduced in the hopping term of Eq.~(\ref{PI-EFKM_eq_H}) 
via the Peierls phase as 
\begin{align}
t^{(\alpha)}_h {\hat c}_{i,\alpha}^{\dag}{\hat c}_{j,\alpha} \; \rightarrow \; t^{(\alpha)}_h 
e^{-i\bm{A}(t)\cdot(\bm{R}_i-\bm{R}_j)}  {\hat c}_{i,\alpha}^{\dag}{\hat c}_{j,\alpha} , 
\label{eq:phase}
\end{align}
where $\bm{R}_j$ is the position of site $j$ and $\bm{A}(t)=A(t)\bm{d}_A$ is the time-dependent 
vector potential along the direction $\bm{d}_A$, thus corresponding to applying the time-dependent electric field 
along $\bm{d}_A$. The velocity of light $c$, 
elementary charge $e$, Planck constant $\hbar$, and the lattice constant are all 
set to 1.  
In this paper, we consider a pump pulse given as 
\begin{align}
A(t) = A_0 e^{-(t-t_0)^2/(2\sigma_p^2)} \cos \left[ \omega_p (t-t_0)  \right] 
\label{A(t)}
\end{align}
with the amplitude $A_0$ and frequency $\omega_p$.  This pulse has a width $\sigma_p$ 
and is centered at time $t_0$ ($>0$) \cite{TIA08,FCNetal12,LSMetal12,HI16,WCMetal17}.

\subsection{Relation to the Hubbard models}\label{sec:relation}

It is well known that the EFKM with $t^{(1)}_h = -t^{(2)}_h$ can be transformed to the repulsive and attractive 
Hubbard models in the pseudospin representation~\cite{Ba02}. 
Here, we summarize the relation among the EFKM with $t^{(1)}_h = -t^{(2)}_h$, the repulsive Hubbard 
model, and the attractive Hubbard model, to emphasize the difference of the condition under which 
the internal SU(2) structure is preserved.

\subsubsection{Repulsive Hubbard model}\label{sec:rhm}

The EFKM with $t^{(2)}_h = -t^{(1)}_h = t_h$ can be transformed into the repulsive Hubbard model by the 
following gauge transformation: 
\begin{align}
\begin{split}
&\hat{c}_{j,1} \to (-1)^j \hat{d}_{j,\uparrow} \\
&\hat{c}_{j,2} \to  \hat{d}_{j,\downarrow}
\end{split}
\label{eq:pht1}
\end{align}
Indeed, the EFKM $\hat{\mathcal{H}}$ is transformed as 
\begin{align}
{\hat {\mathcal{H}}} \to {\hat {\mathcal{H}}}_{\rm R}  = & - t_h\sum_{\langle i, j \rangle,\sigma}  \left( {\hat d}_{i,\sigma}^{\dag}{\hat d}_{j,\sigma} + {\rm H.c.} \right)   
\notag \\
&- \frac{D}{2} \sum_{j}  \left( {\hat n}_{j, \uparrow} -{\hat n}_{j, \downarrow} \right) 
+ U \sum_{j}  {\hat n}_{j, \uparrow}  {\hat n}_{j, \downarrow}  
\label{eq:rhm}                      
\end{align}
provided that the hoppings are finite between sites on different sublattices. 
Here, $\hat{n}_{j,\sigma}={\hat d}_{j,\sigma}^{\dag}{\hat d}_{j,\sigma}$ and $\sigma=\uparrow,\downarrow$. 
${\hat {\mathcal{H}}}_{\rm R}$ is the repulsive Hubbard model in the presence of a Zeeman coupling with a magnetic field $D$.  

Under the transformation in Eq.~(\ref{eq:pht1}), the excitonic (electron-hole) pair operator 
is transformed as 
\begin{align}
\hat{c}_{j,2}^\dagger \hat{c}_{j,1} \to (-1)^j \hat{d}_{j,\downarrow}^\dagger \hat{d}_{j,\uparrow},
\end{align}
thus corresponding to the antiferromagnetic operator 
in the Hubbard model ${\hat {\mathcal{H}}}_{\rm R}$. 
The local $\Delta$-pairing operators are transformed as 
\begin{align}
\begin{split}
&\hat{\Delta}^+_j= \hat{c}^{\dag}_{j,2} \hat{c}^{\dag}_{j,1} \to (-1)^j \hat{d}^{\dag}_{j,\downarrow} \hat{d}^{\dag}_{j,\uparrow}\\
&\hat{\Delta}^z_j=\frac{1}{2}\left( {\hat n}_{j, 1} + {\hat n}_{j, 2} -1 \right) \to 
\frac{1}{2}\left( {\hat n}_{j, \uparrow} + {\hat n}_{j, \downarrow} -1 \right),
\end{split}\label{eq:de}
\end{align}
which correspond to the $\eta$-pairing operators in the Hubbard model ${\hat {\mathcal{H}}}_{\rm R}$, and 
therefore the total $\Delta$-pairing operators, $\hat{\Delta}^\pm$ and $\hat{\Delta}^z$, are transformed to 
the total $\eta$-pairing operators in the Hubbard model ${\hat {\mathcal{H}}}_{\rm R}$ when the model is defined 
on bipartite lattices~\cite{Ya89,EKS91,EFGetal05}. 

It is now clear that the internal SU(2) structure of the EFKM with $t^{(1)}_h = -t^{(2)}_h$ 
in terms of the $\Delta$-pairing operators corresponds to that of the Hubbard model with respect to the 
$\eta$-pairing operators. Here, there are two important remarks. First, this correspondence is true only when 
the model is defined on bipartite lattices. Second, the bipartite condition for lattices 
(and thus $L$ being necessarily even) is required to show 
the internal SU(2) structure of the repulsive Hubbard model in terms of the $\eta$-pairing 
operators~\cite{Ya89,EKS91,EFGetal05}, whereas 
this condition is not assumed to show the internal SU(2) structure of the EFKM with $t^{(1)}_h = -t^{(2)}_h$. 
Therefore, 
in this sense, the model space preserving the internal SU(2) structure is larger for the EFKM than 
the repulsive Hubbard model. 

The same transformation in Eq.~(\ref{eq:pht1}) can transform the hopping term in the presence of the Peierls phase 
in Eq.~(\ref{eq:phase}) as 
\begin{align}
\begin{split}
&\sum_\alpha t^{(\alpha)}_h 
e^{-i\bm{A}(t)\cdot(\bm{R}_i-\bm{R}_j)}  {\hat c}_{i,\alpha}^{\dag}{\hat c}_{j,\alpha}\\
& \quad \quad \quad \rightarrow \; 
t_h  \sum_\sigma e^{-i\bm{A}(t)\cdot(\bm{R}_i-\bm{R}_j)}  {\hat d}_{i,\sigma}^{\dag}{\hat d}_{j,\sigma},
\end{split}
\end{align}
which is exactly the hopping term with the Peierls phase in the Hubbard model. 
Note that here the hoppings are assumed to be finite only between sites on different sublattices. 
Therefore, even the photoinduced dynamics of the EFKM with $t^{(1)}_h = -t^{(2)}_h$ is equivalent to 
the repulsive Hubbard model when the lattice has a bipartite structure. 
Hence, we expect that $\Delta$-pairing is photoinduced in the excitonic insulator of the EFKM, which 
corresponds to the photoinduced $\eta$-pairing in the Mott insulator of the Hubbard model 
found in Ref.~\cite{KSSetal19}.

\subsubsection{Attractive Hubbard model}\label{sec:ahm}

It is also instructive to consider the correspondence between the EFKM and the attractive Hubbard model. 
Since the repulsive Hubbard model and the attractive Hubbard model are mutually transformed via the 
so-called Shiba transformation~\cite{Sh72,EFGetal05}, it is obvious that 
the EFKM with $t^{(2)}_h = -t^{(1)}_h = t_h$  can also be transformed into the attractive 
Hubbard model. For example, the following transformation 
\begin{align}
\begin{split}
&\hat{c}_{j,1} \to \hat{d}_{j,\uparrow}^{\dag} \\
&\hat{c}_{j,2} \to \hat{d}_{j,\downarrow}
\end{split}
\label{eq:pht2}
\end{align}
can transform the EFKM $\hat{\mathcal{H}}$ as 
\begin{align}
{\hat {\mathcal{H}}}  \to  {\hat {\mathcal{H}}}_{\rm A}  =  &- t_h\sum_{\langle i, j \rangle,\sigma}  \left( {\hat d}_{i,\sigma}^{\dag}{\hat d}_{j,\sigma} + {\rm H.c.} \right)   
\notag \\
& +  \frac{D}{2}  \sum_{j}  \left( {\hat n}_{j, \uparrow} + {\hat n}_{j, \downarrow} - 1 \right)  
\notag \\ 
&- U  \sum_{j}  {\hat n}_{j, \uparrow}  {\hat n}_{j, \downarrow}   + U  \sum_j {\hat n}_{j, \downarrow} .                      
\end{align}
We should emphasize that here we do not assume 
the sublattice condition necessary for the transformation from the EFKM to the repulsive Hubbard model 
in Eq.~(\ref{eq:rhm}). 

The same transformation transforms the excitonic pair operator 
as 
\begin{align}
\hat{c}_{j,2}^\dagger \hat{c}_{j,1} \to \hat{d}_{j,\downarrow}^\dagger \hat{d}_{j,\uparrow}^\dagger,
\end{align}
which is the on-site superconducting pair operator 
in the attractive Hubbard model ${\hat {\mathcal{H}}}_{\rm A}$. 
The local $\Delta$-pairing operators are transformed as 
\begin{align}
\begin{split}
&\hat{\Delta}^+_j= \hat{c}^{\dag}_{j,2} \hat{c}^{\dag}_{j,1} \to   \hat{d}^{\dag}_{j,\downarrow} \hat{d}_{j,\uparrow}\\
&\hat{\Delta}^z_j=\frac{1}{2}\left( {\hat n}_{j, 1} + {\hat n}_{j, 2} -1 \right) \to 
- \frac{1}{2}\left( {\hat n}_{j, \uparrow} - {\hat n}_{j, \downarrow}  \right),
\end{split}
\end{align}
corresponding to the spin operators in the attractive Hubbard model ${\hat {\mathcal{H}}}_{\rm A}$, and 
therefore the total $\Delta$-pairing operators, $\hat{\Delta}^\pm$ and $\hat{\Delta}^z$, are transformed to 
the total spin operators in the attractive Hubbard model ${\hat {\mathcal{H}}}_{\rm A}$. 
The internal SU(2) structure of the EFKM with respect to the $\Delta$-pairing operators thus 
corresponds to that of the attractive Hubbard model with respect to the spin operators.  
Note that these correspondences do not require a bipartite lattice structure. 

However, the photoexcited dynamics of the EFKM with $t^{(1)}_h = -t^{(2)}_h$ is different from those 
of the attractive Hubbard model. This is simply because the transformation in Eq.~(\ref{eq:pht2}) transforms 
the hopping term with the Peierls phase in Eq.~(\ref{eq:phase}) as 
\begin{align}
\begin{split}
&\sum_\alpha t^{(\alpha)}_h 
e^{-i\bm{A}(t)\cdot(\bm{R}_i-\bm{R}_j)}  {\hat c}_{i,\alpha}^{\dag}{\hat c}_{j,\alpha}\\
& \quad \quad \quad \rightarrow \; 
t_h  e^{+i\bm{A}(t)\cdot(\bm{R}_i-\bm{R}_j)}  {\hat d}_{i,\uparrow}^{\dag}{\hat d}_{j,\uparrow} \\
& \quad \quad \quad \quad \quad + t_h e^{-i\bm{A}(t)\cdot(\bm{R}_i-\bm{R}_j)}  {\hat d}_{i,\downarrow}^{\dag}{\hat d}_{j,\downarrow} ,
\end{split}
\end{align}
which is different from the hopping term with the Peierls phase in the attractive Hubbard model 
${\hat {\mathcal{H}}}_{\rm A}$. 
The difference of the photoexcited dynamics has been discussed in the context of 
the repulsive and attractive Hubbard models~\cite{KA16}.

\section{Method}\label{sec:method}

In the presence of the external field $\bm{A}(t)$, the Hamiltonian becomes time-dependent, 
$\hat{\mathcal{H}} \rightarrow \hat{\mathcal{H}}(t)$.  To evaluate the state $\ket{\Psi(t)}$ 
under the time-dependent Hamiltonian $\hat{\mathcal{H}}(t)$, we solve the time-dependent 
Schr\"odinger equation numerically 
with the initial condition $\ket{\Psi(t=0)} = \ket{\psi_0}$, where $\ket{\psi_0}$ 
is the ground state of $\hat{\cal{H}}$. 
We employ the time-dependent exact-diagonalization (ED) method based on the 
Lanczos algorithm~\cite{PL86,MA06}.  In this method, the time evolution with a 
short time step $\delta t$ is calculated as 
\begin{align}
\ket{\Psi(t+\delta t)} &\simeq e^{-i \hat{\mathcal{H}}(t) \delta t} \ket{\Psi(t)} 
\notag \\
&\simeq \sum_{\ell=1}^{M_{\rm L}} e^{-i\xi_{\ell} \delta t} \ket{\tilde{\psi}_{\ell}} \braket{\tilde{\psi}_{\ell} |\Psi(t)}, 
\end{align}
where $\xi_{\ell}$ and $\ket{\tilde{\psi}_{\ell}}$ are eigenenergies and eigenvectors 
of $\hat{\mathcal{H}}(t)$, respectively, in the corresponding Krylov subspace 
generated by $M_{\rm L}$ Lanczos iterations~\cite{HI16,PL86,MA06}.  
We use a finite-size cluster of $L$ (even) sites with periodic boundary conditions (PBC).  
We adopt $\delta t = 0.01 / t_h$ and $M_{\rm L} = 15$ for the time evolution, which 
provides results with an almost machine-precision accuracy.  

In order to detect the photoinduced $\Delta$-pairing, we calculate the time evolution 
of the on-site electron-electron pair correlation function defined as  
\begin{align}
P(j,t)  &= \frac{1}{L}\sum_{i}  \bra{\Psi(t)} \left( \hat{\Delta}^{+}_{i+j} \hat{\Delta}^{-}_{i}  + {\rm H.c.} \right) \ket{\Psi(t)} 
\end{align}
and the corresponding structure factor 
\begin{align}
P(\bm{q},t) &=\sum_j e^{i\bm{q} \cdot \bm{R}_j} P(j,t). 
\end{align}
Notice that $P(j,t)$ at $j=0$ is proportional to the double occupancy $n_d (t)$, i.e., 
\begin{align}
P(j \! = \! 0,t)  
= \frac{2}{L}\sum_{i} \bra{\Psi(t)} {\hat n}_{i,1} {\hat n}_{i,2} \ket{\Psi(t)} = 2 n_d (t).
\label{eq:nd}
\end{align} 
Because the ground state of the EFKM has a strong electron-hole pairing correlation,  
we also calculate the excitonic correlation function defined as
\begin{align}
N_X(j,t)  &= \frac{1}{L}\sum_{i}  \bra{\Psi(t)} \left( \hat{b}^{\dag}_{i+j} \hat{b}_{i}  + {\rm H.c.} \right) \ket{\Psi(t)}
\end{align}
and the structure factor 
\begin{align}
N_X(\bm{q},t) &=\sum_j e^{i\bm{q} \cdot \bm{R}_j} N_X(j,t) ,
\end{align}
where $b^{\dag}_j = c^{\dag}_{j,2}c_{j,1}$ is the creation operator of an exciton.  

Hereafter, we define $t_h \equiv |t^{(1)}_h|$ and use $t_{h}$  ($t_{h}^{-1}$) as the unit of energy (time).  
We set the total number of electrons $N=N_1+N_2$ to be $L$, i.e., half filling. 
Note that the number $N_{\alpha}$ of electrons for each orbital $\alpha\,(=1,2)$ is conserved even in the 
presence of the external field in Eq.~(\ref{eq:phase}). 
$N_{\alpha}$ depends on the values of $D$ and $U$. The results shown in the next section 
are for $t^{(2)}_h = -t^{(1)}_h >0$ and $D>0$ with $N_1 > N_2$.

\section{Numerical results}\label{sec:results}

The correspondence shown in Sec.~\ref{sec:rhm} implies that $\Delta$-pairing can be photoinduced in the excitonic 
insulator of the EFKM with the direct-gap-type band structure (i.e., $t^{(1)}_h =- t^{(2)}_h$). 
Because the $\eta$-pairing in the Hubbard model studied previously in Ref.~\cite{KSSetal19} corresponds to 
the EFKM with $t^{(1)}_h = -t^{(2)}_h$ and $D=0$, here we focus on the case with $D\ne0$ as well as a 
nonbipartite lattice. The photoinduced interband $\eta$-pairing in the EFKM with the indirect-gap-type band 
structure (i.e., $t^{(1)}_h = t^{(2)}_h$) is discussed in Appendix~\ref{sec:app-eta}. 

\subsection{1D system}\label{sec:1d}

\begin{figure}[t]
\begin{center}
  \includegraphics[width=0.9\columnwidth]{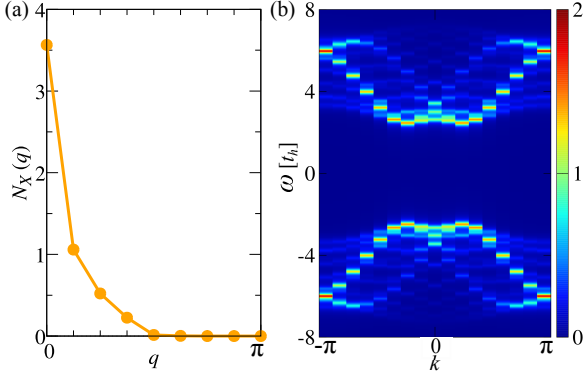}
\caption{
(a) Excitonic (i.e., electron-hole pair) structure factor $N_X(q)$ and (b) single-particle excitation spectrum 
$A(k,\omega)$ for the ground state of the 1D EFKM with $t_h^{(2)}=-t_h^{(1)}=t_h$, $U=8t_h$, and $D=0.75t_h$ 
in $L=16$, where  $N_1=12$ and $N_2 =4$. The Fermi energy is set at $\omega=0$ in (b). 
Here, the single-particle excitation spectrum is defined as 
$A(k,\omega)=\sum_\alpha\langle\psi_0|\hat c_{k,\alpha}^\dag\delta_{\gamma}(\omega+\hat{\cal H}-E_0)\hat c_{k,\alpha}|\psi_0\rangle + \sum_\alpha\langle\psi_0|\hat c_{k,\alpha}\delta_{\gamma}(\omega-\hat{\cal H}+E_0)\hat c_{k,\alpha}^\dag|\psi_0\rangle$, where $\hat c_{k,\alpha}^\dag$ is the Fourier transform of $\hat c_{j,\alpha}^\dag$ [also see Eq.~(\ref{eq:x_jj})].  
The broadening factor $\gamma$ in $A(k,\omega)$ is $0.1 t_h$. 
}
\label{fig:efkm_initial}
\end{center}
\end{figure}

First, we show the results for the 1D EFKM with $t^{(2)}_h = -t^{(1)}_h = t_h$.  
Here, we set the vector potential $\bm{A}(t)=A(t)\bm{e}_x$ along the chain direction, i.e. 
$t^{(\alpha)}_h {\hat c}_{j,\alpha}^{\dag}{\hat c}_{j+1,\alpha} \rightarrow t^{(\alpha)}_h e^{iA(t)} {\hat c}_{j,\alpha}^{\dag}{\hat c}_{j+1,\alpha}$. 
We assume that $U=8t_h$ and $D=0.75t_h$ in 
$L=16$, for which the ground state of the 1D EFKM, i.e., the initial state before the pulse irradiation, 
is the excitonic insulator   
with $N_1=12$ and $N_2 =4$ 
(see Fig.~\ref{fig:efkm_initial}).

\begin{figure}[t]
\begin{center}
  \includegraphics[width=\columnwidth]{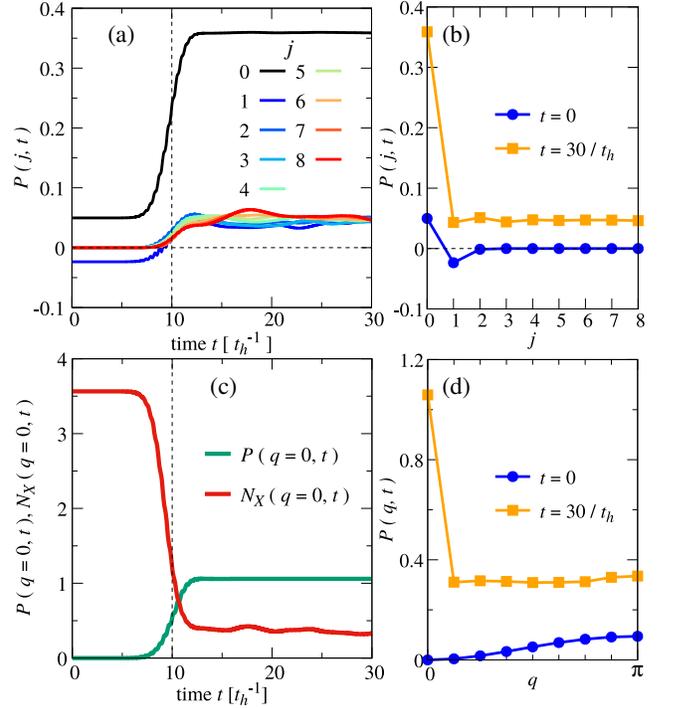}
\caption{
(a) Time evolution of the on-site electron-electron pair correlation function $P(j,t)$.  
(b) $P(j,t)$ at $t=0$ (blue circles) and $t=30/t_h$ (orange squares). 
(c) Time evolution of the electron-electron pair structure factor $P(q,t)$ and the excitonic 
(i.e., electron-hole pair) structure factor $N_X(q,t)$ at $q=0$. 
(d) $P(q,t)$ at $t=0$ (blue circles) and $t=30/t_h$ (orange squares).  
The results are for the 1D EFKM with $t^{(2)}_h = -t^{(1)}_h = t_h$, $U=8t_h$, and $D=0.75t_h$ in 
$L=16$. In this case, the initial state before the pulse irradiation (i.e., the ground state of the 1D EFKM) 
has $N_1=12$ and $N_2 =4$. 
We set $A_0=0.4$, $\omega_p=7t_h$, $\sigma_p=2/t_h$, and $t_0=10/t_h$ for $A(t)$.  
The vertical dashed lines in (a) and (c) indicate $t_0$. 
}\label{fig:pair_corr}
\end{center}
\end{figure}

Figure~\ref{fig:pair_corr}(a) shows the time evolution of the real-space electron-electron pair correlation function $P(j,t)$.
We confirm the enhancement of $P(j,t)$ at $j=0$, corresponding to $n_d (t)$, by the pulse irradiation, which is similar to 
the case in the Hubbard model~\cite{KSSetal19}. 
As we expected, the electron-electron pair correlation $P(j \! \ne \! 0,t)$ is also enhanced by the pulse irradiation 
and becomes positive for all sites. 
As shown in Fig.~\ref{fig:pair_corr}(b), the pair correlation after the pulse irradiation extends to longer distances 
over the cluster, while the pair correlation is essentially absent in the initial excitonic insulating state before the pulse 
irradiation. 
It is also clear that the sign of $P(j,t)$ is positive for all sites, 
and consequently the pair structure factor $P(q,t)$ shows a sharp peak at $q=0$ [see Fig.~\ref{fig:pair_corr}(d)]. 
The time evolution of $P(q,t)$ and the excitonic structure factor $N_X(q,t)$ are also calculated at $q=0$ 
in Fig.~\ref{fig:pair_corr}(c). 
The excitonic correlation $N_X(q=0,t)$ is indeed large in the initial state, 
as shown also in Fig.~\ref{fig:efkm_initial}(a), and is significantly suppressed 
by the pulse irradiation. In contrast,  
the pair correlation $P(q=0,t)$ is strongly enhanced 
despite that it is exactly zero before the pulse irradiation.

\begin{figure}[t]
\begin{center}
\includegraphics[width=1.0\columnwidth]{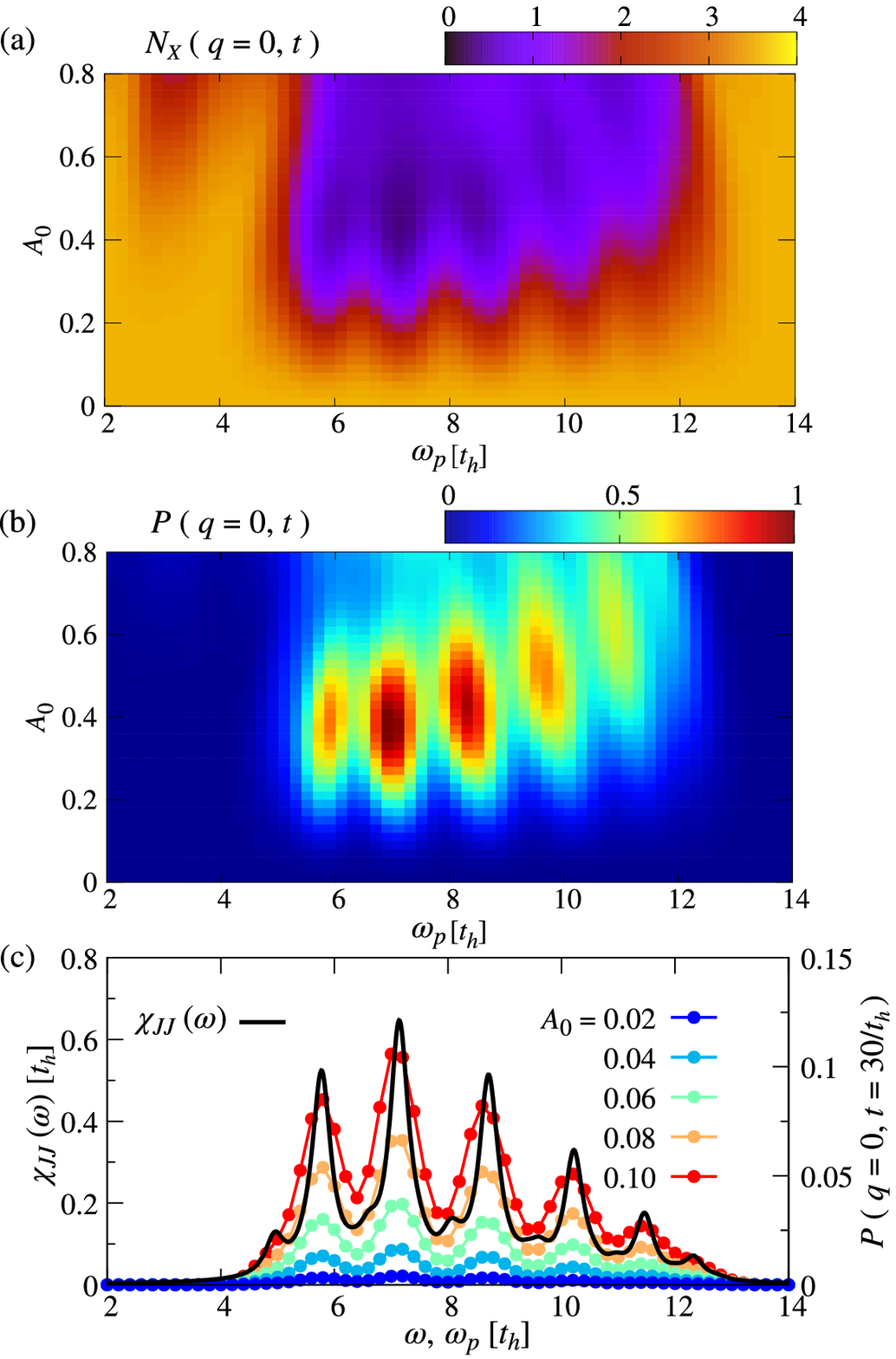}
\caption{
Contour plots of 
(a) the excitonic structure factor $N_X(q=0,t)$ 
averaged from $t=20/t_h$ to $40/t_h$ and 
(b) the electron-electron pair structure factor $P(q=0,t)$ at $t=30/t_h$ in the parameter space of $\omega_p$ and $A_0$.  
(c) Optical spectrum $\chi_{JJ}(\omega)$ calculated for the ground state of the EFKM, which is compared with 
$P(q=0,t=30/t_h)$ as a function of $\omega_p$ for different values of $A_0$.  
The results are for the 1D EFKM with  $t^{(2)}_h = -t^{(1)}_h = t_h$, $U=8t_h$, and $D=0.75t_h$ in 
$L=16$. In this case, the initial state before the pulse irradiation (i.e., the ground state of the 1D EFKM) 
has $N_1=12$ and $N_2 =4$. 
We set $\sigma_p=2/t_h$ and $t_0=10/t_h$ for $A(t)$.  
The broadening factor $\gamma$ in $\chi_{JJ}(\omega)$ is $0.2 t_h$ in (c). 
}\label{fig:optimal}
\end{center}
\end{figure}

In order to identify the optimal control parameters for the enhancement of $P(q=0,t)$, Fig.~\ref{fig:optimal}(b) 
shows the contour plot of $P(q=0,t)$ 
after the pulse irradiation with different values of $A_0$ and $\omega_p$.
As shown in Fig.~\ref{fig:optimal}(c), for small $A_0$, we find that the peak structure of $P(q=0,t)$ 
as a function of $\omega_p$ are essentially 
the same  as the ground-state optical spectrum  
\begin{align}
\chi_{JJ}(\omega) & = \frac{1}{L}\bra{\psi_0} \hat{J} \delta_{\gamma}(\omega-\hat{\mathcal{H}}+E_0) \hat{J} \ket{\psi_0}      
\notag \\
  = & -\frac{1}{\pi L} {\rm Im} \left[ \bra{\psi_0} \hat{J} \frac{1}{\omega-\hat{\mathcal{H}}+E_0+i\gamma} \hat{J} \ket{\psi_0}  \right], 
  \label{eq:x_jj}
\end{align}
where $\ket{\psi_0}$ is the ground state of $\hat{\cal{H}}$ with its energy $E_0$,  
\begin{align}
\hat{J}=i\sum_{j,\alpha}t_{h}^{(\alpha)} (\hat{c}^{\dag}_{j+1,\alpha} \hat{c}_{j,\alpha} - \hat{c}^{\dag}_{j,\alpha} \hat{c}_{j+1,\alpha})
\end{align}
is the current operator, and $\gamma$ is the broadening factor~\cite{LGBetal14,HI17}.
As discussed later, this can be understood on the basis of the internal SU(2) structure of the EFKM 
with $t^{(1)}_h = -t^{(2)}_h$. 
We also notice in Fig.~\ref{fig:optimal}(b) that with further increasing $A_0$, where the nonlinearity becomes important, 
the peak structure of $P(q=0,t)$ as a 
function of $\omega_p$ slightly shifts from that of $\chi_{JJ}(\omega)$. 
The optimal parameters for the enhancement of $P(q=0,t)$ is $\omega_p\sim7t_h$ and $A_0\sim0.4$ for the system studied in Fig.~\ref{fig:optimal}. 
On the other hand, as shown in Fig.~\ref{fig:optimal}(a), the 
excitonic correlation is strongly suppressed in the region where the electron-electron pair correlation is enhanced. 
We should emphasize that the enhancement of $P(q=0,t)$ cannot be simply explained by a 
dynamical phase transition induced by effectively varying the model parameters 
because there is no region in the ground state phase diagram 
of the EFKM~\cite{EKOetal14}, showing large electron-electron pairing correlations.

Two remarks are in order. First, 
the spike structure of $P(q=0,t)$ found in Fig.~\ref{fig:optimal}(b) depends on the system size and 
is expected to be smooth in the thermodynamic limit ($L\rightarrow \infty$), as in the case for the 
optical spectrum $\chi_{JJ}(\omega)$, shown in Fig.~\ref{fig:optimal}(c), where the spike structure 
becomes less pronounced and eventually smooth with increasing $L$~\cite{FMSetal91,JGE00}. 
Second, the electron-electron pair structure factor $P(q=0,t)$ is most apparently enhanced in the frequency region  
of $5t_h\alt\omega_p\alt 12t_h$, which corresponds approximately to the single-particle excitation gaps at different 
momenta for the initial state [see Fig.~\ref{fig:efkm_initial}(b)]. 

\begin{figure}[t]
\begin{center}
\includegraphics[width=\columnwidth]{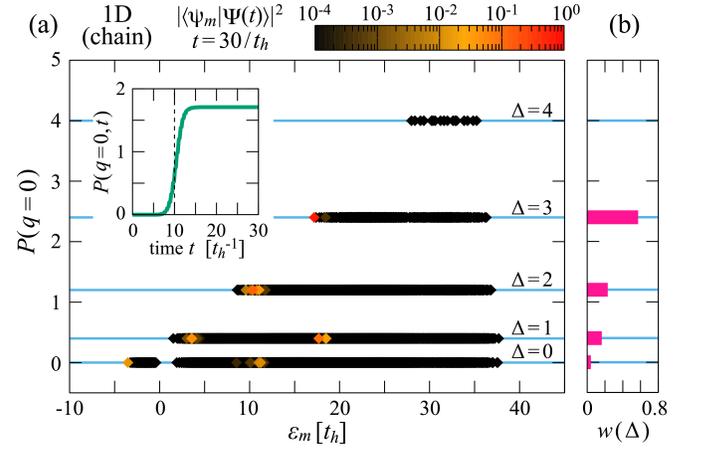}
\caption{
(a)~All the eigenenergies $\varepsilon_m$ and $P(q=0)$ for the eigenstates $\ket{\psi_m}$ of the half-filled 1D EFKM 
$\hat{\cal{H}}$ with  $t^{(2)}_h = -t^{(1)}_h = t_h$ for $L=10$ under PBC at $U=8t_h$ and $D=0.4t_h$, 
where $N_1= 6$ and $N_2 =4$. 
The color of each point (diamond) indicates the weight $|\braket{\psi_m|\Psi(t)}|^2$ of the eigenstate $\ket{\psi_m}$ in the photoinduced state $\ket{\Psi(t)}$ at $t=30/t_h$ 
for $A(t)$ with $A_0=0.3$, $\omega_p=7t_h$, $\sigma_p=2/t_h$, and  $t_0=10/t_h$.
When the eigenstates are degenerate, 
the color indicates the sum of $|\braket{\psi_m|\Psi(t)}|^2$ over these degenerate states.  
The inset shows the time evolution of $P(q=0,t)$ 
for $\ket{\Psi(t)}$. 
(b) The total weight $w(\Delta)$ of $|\braket{\psi_m|\Psi(t)}|^2$ over the eigenstates $\ket{\psi_m}$ 
with the same value of $\Delta$ in (a). 
Note that $\sum_{\Delta}w(\Delta)=1$. 
}
\label{fig:structure}
\end{center}
\end{figure}

To understand the origin of the enhancement of the on-site electron-electron pair correlations by the pulse irradiation, 
let us now elucidate the nature of the photoinduced state $\ket{\Psi(t)}$ in terms of the $\Delta$ pairs. 
For this purpose, 
we calculate the eigenenergies $\varepsilon_m$ and the electron-electron pair structure factors $P(q\!=\! 0)$ 
for all the eigenstates $\ket{\psi_m}$ of the 1D EFKM $\hat{\cal{H}}$ at half filling. 
As shown in Fig.~\ref{fig:structure}(a), the structure factor $P(q\!= 0)$ for each eigenstate is exactly quantized. 
This is understood because each eigenstate $\ket{\psi_m}$ of $\hat{\cal{H}}$ is also the eigenstate of 
$\Delta$-pairing operators $\hat{\Delta}^2$ and $\hat{\Delta}_z$ with the eigenvalues $\Delta$ and $\Delta_z$, 
respectively [see Eqs.~(\ref{eq:cr_delta}) and (\ref{eq:cr_h})]. 
Therefore, the structure factor $P(q\!=\! 0)$ is given as
\begin{align}
P(q=0) &= \frac{2}{L} \bra{\psi_{m}}\hat{\Delta}^+\hat{\Delta}^- \ket{\psi_{m}} \nonumber\\
&= \frac{2}{L} \bra{\psi_{m}}\left(\hat{\Delta}^2 - \hat{\Delta}_z^2 + \hat{\Delta}_z \right)\ket{\psi_{m}} \nonumber\\
&= \frac{2}{L} \Delta(\Delta +1)
\label{eq:Pq=0}
\end{align}
with 
$\Delta=0,1,\cdots, N_2$, where $N_2$ ($\le N_1$)
is the maximum number of $\Delta$ pairs 
and   we have used $\Delta_z=0$ at half filling. Thus, the quantized value corresponds to the eigenvalue $\Delta$ 
of $\hat{\Delta}^2$ for the eigenstate $\ket{\psi_m}$ of $\hat{\cal{H}}$.

We can construct the eigenstate with the number $N_\Delta$ of $\Delta$ pairs from the LWS for the 
$\Delta$-pairing operators 
as 
\begin{align}
|\psi_{N_\Delta}\rangle \! \propto \! \left(\hat{\Delta}^+\right)^{N_\Delta} \!
\left| \Delta \!=\! \frac{L}{2} \!-\! \frac{N_1 \!+\! N_2 \!-\! 2N_\Delta}{2}, \Delta_z \!=\! -\Delta \right\rangle,
\end{align}
where we assume that there are $N_1$ and $N_2$ electrons for orbitals 1 and 2, respectively, in $|\psi_{N_\Delta}\rangle$,  
and $L\ge N_1+N_2-2N_\Delta$. 
Since we are at half filling, i.e., $N_1+N_2=L$, $|\psi_{N_\Delta}\rangle\propto  \left(\hat{\Delta}^+\right)^{N_\Delta} 
|\Delta=N_\Delta,\Delta_z=-\Delta\rangle \propto |\Delta=N_\Delta,\Delta_z=0\rangle$. Therefore, in this case, 
$\bra{\psi_{N_\Delta}}\hat{\Delta}^+\hat{\Delta}^- \ket{\psi_{N_\Delta}}=N_\Delta(N_\Delta+1)$ and 
thus $P(q=0)= 2 N_\Delta(N_\Delta+1)/L$. 
Comparing with Eq.~(\ref{eq:Pq=0}), we can thus 
notice that the eigenvalue $\Delta$ of $\hat{\Delta}^2$ for $\ket{\psi_{m}}$ corresponds to the number 
$N_\Delta$ of $\Delta$ pairs contained in $\ket{\psi_{m}}$ at half filling.

\begin{figure}[t]
\begin{center}
\includegraphics[width=0.93\columnwidth]{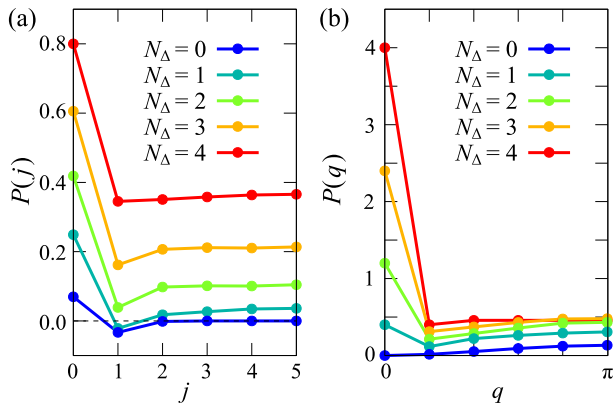}
\caption{
(a) On-site electron-electron pair correlation function $P(j)$ and (b) the corresponding structure factor $P(q)$ 
for the half-filled 
eigenstate $\ket{\psi_{N_{\Delta}}}$ with the different number $N_\Delta$ of $\Delta$ pairs. 
The eigenstate $\ket{\psi_{N_{\Delta}}}$ is constructed from the exact ground state 
of the 1D EFKM with $N_1-N_\Delta$ and $N_2-N_\Delta$ electrons for orbitals 1 and 2, respectively, 
calculated by the ED method, for $U=8t_h$ and $D=0.4 t_h$ in $L=10$ under PBC, 
where $N_1= 6$ and $N_2 =4$. 
}
\label{fig:structure2}
\end{center}
\end{figure}

As an example, we construct $|\psi_{N_\Delta}\rangle$ from the exact ground state 
$|\psi^{\rm (GS)}_{N_1-N_\Delta,N_2-N_\Delta}\rangle$ of the 1D EFKM with $N_1-N_\Delta$ 
($N_2-N_\Delta$) electrons for orbital 1 (2), which is the LWS for the $\Delta$-pairing operators. 
Figure~\ref{fig:structure2} shows the on-site electron-electron pair correlations $P(j)$ 
and the corresponding structure factor $P(q)$ for $|\psi_{N_\Delta}\rangle$ containing different number 
$N_\Delta$ of $\Delta$ pairs. 
With increasing $N_\Delta$, the enhancement of $P(j)$ and $P(q\!=\! 0)$ are clearly observed. 
Their structures are in good qualitative agreement with the electron-electron pair correlations of 
the photoinduced state $\ket{\Psi(t)}$ shown in Figs.~\ref{fig:pair_corr}(b) and \ref{fig:pair_corr}(d). 
Notice that the quantized values of $P(q\!=\! 0)$ found in Fig.~\ref{fig:structure}(a) corresponds exactly to the 
values of $P(q\!=\! 0)$ in Fig.~\ref{fig:structure2}(b).

In Fig.~\ref{fig:structure}(a), the color of each point indicates the weight $|\braket{\psi_m|\Psi(t)}|^2$ of the 
eigenstate $\ket{\psi_m}$ in the photoinduced state $\ket{\Psi(t)}$
that exhibits the strong enhancement of $P(q\!=\! 0,t)$ after the pulse irradiation [see the inset of Fig~\ref{fig:structure}(a)].
We find that the state $\ket{\Psi(t)}$ after the pulse irradiation contains the nonzero weights of the eigenstates 
$\ket{\psi_m}$ with finite $\Delta$ [also see Fig.~\ref{fig:structure}(b)]. 
This is precisely the reason for the photoinduced enhancement of $P(q \! = \! 0,t)$. 
The EFKM itself has the eigenstates with $P(q\!=\! 0)\ne0$ and the photoinduced state 
$\ket{\Psi(t)}$ captures the weights of those eigenstates. Since the number $N_\Delta$ of $\Delta$ pairs in $\ket{\psi_m}$ 
is $\Delta$, the photoinduced state $\ket{\Psi(t)}$ contains a finite number of $\Delta$ pairs.

The process of the enhancement of $P(q \! = \! 0,t)$ is essentially the same as the photoinduced $\eta$-pairing 
in the Hubbard model~\cite{KSSetal19} and is understood as follows. 
Before the pulse irradiation, the initial state is the ground state of the EFKM 
$\hat{\cal H}$ with $\ket{\Delta=0,\Delta_z=0}$, i.e., the singlet state for the $\Delta$-pairing operators, and $P(q=0) = 0$. 
The pulse irradiation via $A(t)$ breaks the commutation relation as
$[\hat{\mathcal{H}}(t),\hat{\Delta}^+] = [\hat{\mathcal{H}},\hat{\Delta}^+] + \sum_{k} F(k,t) \hat{c}^{\dag}_{-k,2} \hat{c}^{\dag}_{k ,1}$ with $F(k,t) = 4t^{(1)}_h \sin[A(t)] \sin k$ for $t^{(1)}_h=-t^{(2)}_h$,
and this transient breaking of the internal SU(2) structure stirs states with different values of $\Delta$. 
After the pulse irradiation, the Hamiltonian again satisfies the commutation relation because $A(t) = 0$
but  the state $\ket{\Psi(t)}$ now contains components of $\ket{\Delta\ne0,\Delta_z=0}$, which enhance $P(q=0,t)$.

However, this does not explain details of the spectrum structure in Fig.~\ref{fig:structure}(a), i.e., why 
some particular eigenstates $\ket{\psi_m}$ are selectively excited in the 
photoinduced state $\ket{\Psi(t)}$ and others are not. For example, focusing the eigenstates $\ket{\psi_m}$ with the 
eigenenergies $\varepsilon_m\sim 10 t_h$, the eigenstates with $\Delta=0$ and $2$ have large overlap 
$|\braket{\psi_m|\Psi(t)}|^2$ with the photoinduced state $\ket{\Psi(t)}$, but no overlap with 
the eigenstates with $\Delta=1$ is observed in this eigenenergy region. 
As shown in Sec.~\ref{sec:sr}, the understanding of the detailed spectrum structure requires the symmetry argument 
based on the internal SU(2) structure of the EFKM with respect to the $\Delta$-pairing operators.

\subsection{Two dimensional systems}

\subsubsection{Square lattice}\label{sec:sq}

Similarly, $\Delta$ pairs can be photoinduced in the two dimensional (2D) EFKM in the square lattice. 
This is expected because, as described in Sec.~\ref{sec:rhm}, when the system is bipartite, the 2D EFKM 
with $t^{(1)}_h = -t^{(2)}_h$ can be mapped onto the repulsive Hubbard model where $\eta$ pairs can be 
induced by the pulse irradiation~\cite{KSSetal19}. Since the $\eta$ pair in the repulsive Hubbard model 
corresponds to the $\Delta$ pair in the EFKM with $t^{(1)}_h = -t^{(2)}_h$ [see Eq.~(\ref{eq:de})], 
the photoinduced $\Delta$ pairs are anticipated in the EFKM with $t^{(1)}_h = -t^{(2)}_h$ when the system 
is bipartite. 

\begin{figure}[t]
\begin{center}
\includegraphics[width=1.0\columnwidth]{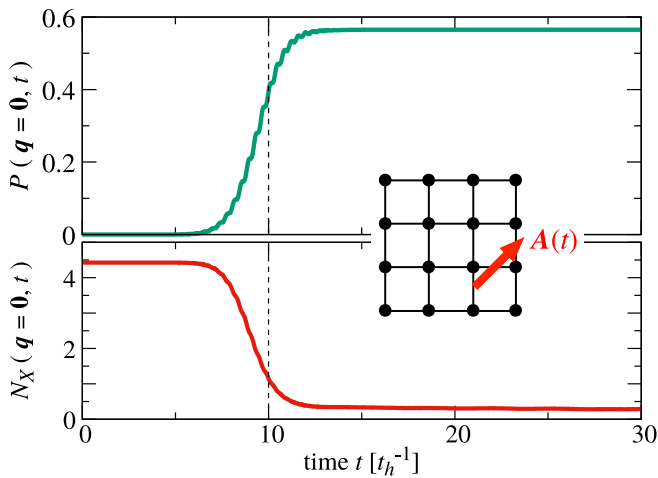}
\caption{
Time evolution of the electron-electron pair structure factor $P(\bm{q},t)$ and the excitonic 
(i.e., electron-hole pair) structure factor $N_X(\bm{q},t)$ at $\bm{q}=\bm{0}=(0,0)$ 
for the 2D EFKM with $t^{(2)}_h = -t^{(1)}_h = t_h$, $U=8t_h$, and $D=t_h$ in 
a $4\times4$ square lattice under PBC. In this case, the initial state before the pulse irradiation 
(i.e., the ground state of the 2D EFKM) 
has $N_1=12$ and $N_2 =4$. 
The time-dependent vector potential $\bm{A}(t) = A(t)(\bm{e}_x+\bm{e}_y)$ is applied 
along the diagonal direction (indicated in the figure).
We set $A_0=0.4$, $\omega_p=8t_h$, $\sigma_p=2/t_h$, and $t_0=10/t_h$ for $A(t)$. 
The vertical dashed line indicates $t_0$. 
}
\label{fig:2d_corr}
\end{center}
\end{figure}

Figure~\ref{fig:2d_corr} shows the time evolution of the electron-electron pair structure factor $P(\bm{q},t)$ 
and the excitonic (electron-hole) pair structure factor $N_X(\bm{q},t)$ at $\bm{q}=\bm{0}=(0,0)$ for the 
2D EFKM with $t^{(1)}_h = -t^{(2)}_h$ on a $4 \times 4$ cluster with PBC. 
Here, the time-dependent vector potential $\bm{A}(t)$ is applied along the diagonal direction, i.e., 
$\bm{A}(t) = A(t)(\bm{e}_x+\bm{e}_y)$, where $\bm{e}_{x(y)}$ is the unit vector along the $x$ ($y$) direction 
and $A(t)$ is defined in Eq.~(\ref{A(t)}). 
As in the 1D case shown in Fig.~\ref{fig:pair_corr}(c), the initial ground state is the excitonic insulator and the 
excitonic correlation $N_X(\bm{q}=\bm{0},t)$ is significantly suppressed after the pulse irradiation, while 
the enhancement of the on-site electron-electron pairing correlation $P(\bm{q}=\bm{0},t)$ 
by the pulse irradiation is indeed observed. 

\subsubsection{Triangular lattice}

A nontrivial system is the 2D EFKM in the triangular lattice, for which there is no correspondence 
to the repulsive Hubbard model, as discussed in Sec.~\ref{sec:rhm}. 
In contrast to the case of the $\eta$-pairing operators in the Hubbard model, 
the $\Delta$-pairing operators in the EFKM satisfy $[\hat{\mathcal{H}},\hat{\Delta}^{\pm}] = \pm U \hat{\Delta}^{\pm}$, 
regardless of whether the lattice is bipartite or nonbipartite, 
since $\epsilon_2(-\bm{k})=-\epsilon_1(\bm{k})$ when $t^{(1)}_h=-t^{(2)}_h$ 
[see Eq.~(\ref{PI-EFKM_eq_H-Delta})]. Therefore, the internal SU(2) structure with respect to the 
$\Delta$-pairing operators are preserved for the 2D EFKM with $t^{(1)}_h=-t^{(2)}_h$ in the triangular lattice. 
This implies that the similar results found for the 1D EFKM in Sec.~\ref{sec:1d} and for the square EFKM 
in Sec.~\ref{sec:sq} are expected for the triangular EFKM.

\begin{figure}[t]
\begin{center}
\includegraphics[width=\columnwidth]{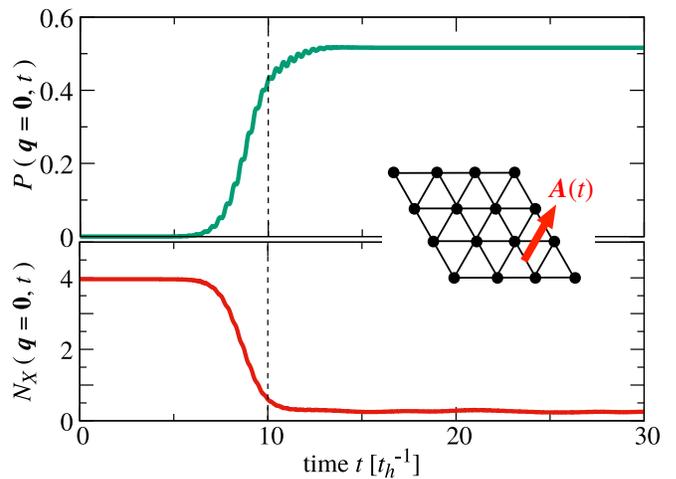}
\caption{
Time evolution of the electron-electron pair structure factor $P(\bm{q},t)$ and the excitonic 
(i.e., electron-hole pair) structure factor $N_X(\bm{q},t)$ at $\bm{q}=\bm{0}=(0,0)$ 
for the 2D EFKM with $t^{(2)}_h = -t^{(1)}_h = t_h$, $U=8t_h$, and $D=1.3t_h$ in 
a $4\times4$ triangular lattice under PBC. In this case, the initial state before the pulse irradiation 
(i.e., the ground state of the 2D EFKM) 
has $N_1=12$ and $N_2 =4$. 
The time-dependent vector potential $\bm{A}(t) = A(t)(\frac{1}{2}\bm{e}_x+\frac{\sqrt{3}}{2}\bm{e}_y)$ is applied 
along the direction indicated in the figure. 
We set $A_0=0.6$, $\omega_p=8t_h$, $\sigma_p=2/t_h$, and $t_0=10/t_h$ for $A(t)$. 
The vertical dashed line indicates $t_0$. 
}
\label{fig:2d_corr_t}
\end{center}
\end{figure}

Figure~\ref{fig:2d_corr_t} shows the time evolution of the electron-electron pair structure factor $P(\bm{q},t)$ 
and the excitonic (electron-hole) pair structure factor $N_X(\bm{q},t)$ at $\bm{q}=\bm{0}$ for the 
2D EFKM with $t^{(1)}_h = -t^{(2)}_h$ on a $4 \times 4$ triangular cluster with PBC. 
Here, the time-dependent vector potential $\bm{A}(t) = A(t)(\frac{1}{2}\bm{e}_x+\frac{\sqrt{3}}{2}\bm{e}_y)$ 
is applied in the direction indicated in Fig.~\ref{fig:2d_corr_t}. 
As in the square lattice, we find that the excitonic correlation $N_X(\bm{q}=\bm{0},t)$ is suppressed by the 
pulse irradiation, while the on-site electron-electron pairing correlation $P(\bm{q}=\bm{0},t)$ is enhanced. 

\begin{figure}[t]
\begin{center}
\includegraphics[width=\columnwidth]{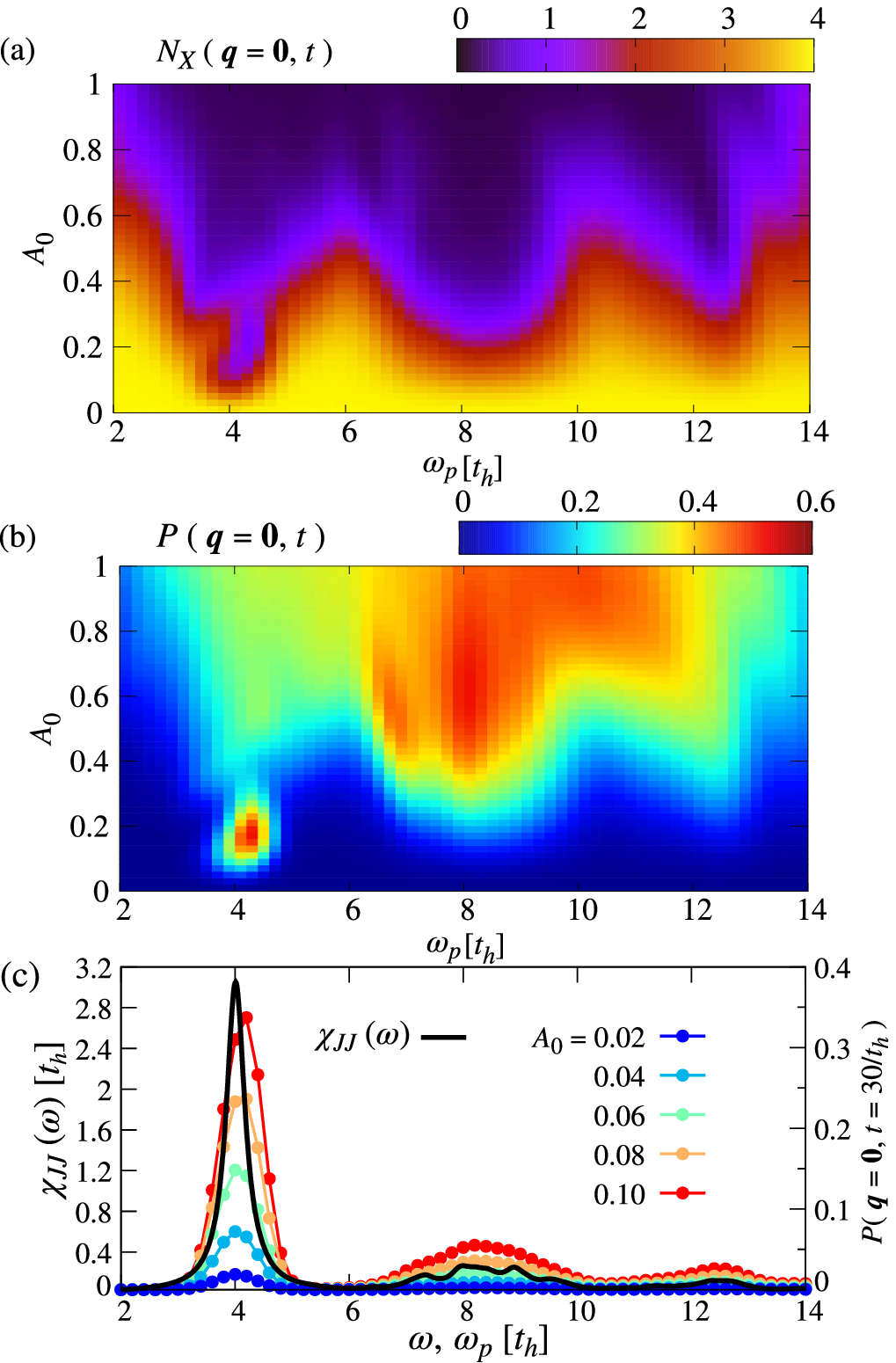}
\caption{
Contour plots of 
(a) the excitonic structure factor $N_X(\bm{q}=\bm{0},t)$ averaged 
from $t=20/t_h$ to $40/t_h$ and 
(b) the electron-electron pair structure factor $P(\bm{q}=\bm{0},t)$ at $t=30/t_h$ in the parameter space of 
$\omega_p$ and $A_0$.  
(c) Optical spectrum $\chi_{JJ}(\omega)$ calculated for the ground state of the EFKM, which is compared with 
$P(\bm{q}=\bm{0},t=30/t_h)$ as a function of $\omega_p$ for different values of $A_0$.  
The results are for the 2D EFKM with $t^{(2)}_h = -t^{(1)}_h = t_h$, $U=8t_h$, and $D=1.3t_h$ in 
a $4\times4$ triangular lattice under PBC. In this case, the initial state before the pulse irradiation 
(i.e., the ground state of the 2D EFKM) has $N_1=12$ and $N_2 =4$. 
We set $\sigma_p=2/t_h$ and $t_0=10/t_h$ for $A(t)$.  
The broadening factor $\gamma$ in $\chi_{JJ}(\omega)$ is $0.2t_h$ in (c).
}\label{fig:optimal_tr}
\end{center}
\end{figure}

Figures~\ref{fig:optimal_tr}(a) and \ref{fig:optimal_tr}(b) show the results of the optimal parameter 
$A_0$ and $\omega_p$ search for the enhancement of the on-site electron-electron pair correlation 
in the photoexcited state. As in the case for the 1D EFKM shown in Figs.~\ref{fig:optimal}(a) and \ref{fig:optimal}(b), 
the electron-electron pair correlation is most efficiently enhanced when the excitonic electron-hole pair 
correlation is most significantly suppressed. 
We also find in Fig.~\ref{fig:optimal_tr}(c) that 
the electron-electron pair correlation factor $P(\bm{q}=\bm{0},t=30/t_h)$ as a function of $\omega_p$ is essentially 
the same, when $A_0$ is small, as the optical spectrum $\chi_{JJ}(\omega)$ calculated for 
the ground state of the 2D EFKM in the triangular lattice. As discussed in Sec.~\ref{sec:sr}, this is due to the 
symmetry property of the current operator $\hat{J}$ with respect to the $\Delta$-pairing operators. 
 
\begin{figure}[t]
\begin{center}
\includegraphics[width=\columnwidth]{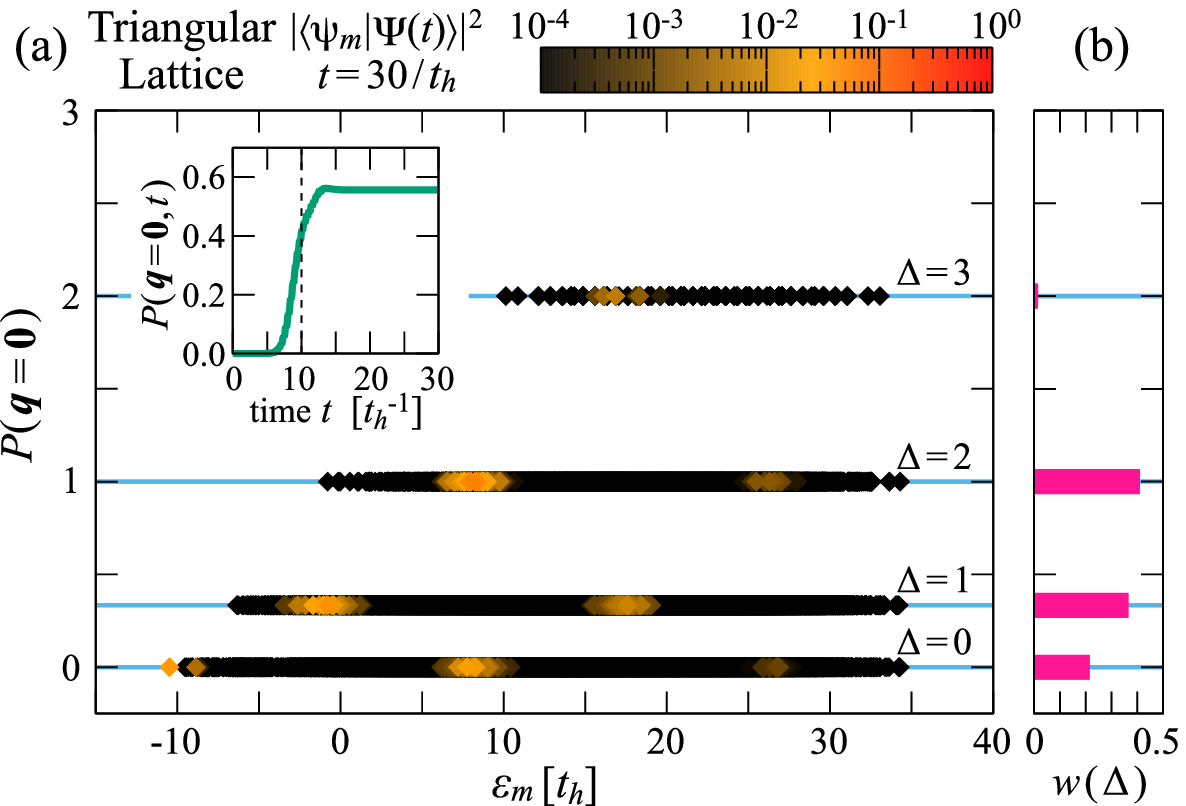}
\caption{
(a)~All the eigenenergies $\varepsilon_m$ and $P(\bm{q}=\bm{0})$ for the eigenstates $\ket{\psi_m}$ 
of the half-filled 2D EFKM $\hat{\cal{H}}$ with $t^{(2)}_h = -t^{(1)}_h = t_h$ 
in a $4\times3$ triangular cluster under PBC at $U=8t_h$ and $D=0.8t_h$, 
where $N_1= 9$ and $N_2 =3$. 
The color of each point (diamond) indicates the weight $|\braket{\psi_m|\Psi(t)}|^2$ of the eigenstate $\ket{\psi_m}$ in the photoinduced state $\ket{\Psi(t)}$ at $t=30/t_h$ 
for $A(t)$ with $A_0=0.4$, $\omega_p=9t_h$, $\sigma_p=2/t_h$, and  $t_0=10/t_h$.
When the eigenstates are degenerate, 
the color indicates the sum of $|\braket{\psi_m|\Psi(t)}|^2$ over these degenerate states.  
The inset shows the time evolution of $P(\bm{q}=\bm{0},t)$ for $\ket{\Psi(t)}$. 
(b) The total weight $w(\Delta)$ of $|\braket{\psi_m|\Psi(t)}|^2$ over the eigenstates $\ket{\psi_m}$ 
with the same value of $\Delta$ in (a). 
Note that $\Delta$ corresponds to the number $N_\Delta$ of $\Delta$ pairs at half filling and  
$\sum_{\Delta}w(\Delta)=1$. 
}
\label{fig:structure_tr}
\end{center}
\end{figure}

To better understand the nature of the photoexcited state $|\Psi(t)\rangle$, we calculate the electron-electron pair structure 
factor $P(\bm{q})$ at $\bm{q}=\bm{0}$ for all the eigenstates $\ket{\psi_m}$ of the 2D EFKM $\hat{\cal{H}}$ in the 
triangular lattice. 
As shown in Fig.~\ref{fig:structure_tr}(a), we find that $P(\bm{q}=\bm{0})$ is exactly quantized for all the eigenstates 
$\ket{\psi_m}$ and the quantized values are give in Eq.~(\ref{eq:Pq=0}). This is because any eigenstate 
$\ket{\psi_m}$ of the 2D EFKM $\hat{\cal{H}}$ in the triangular lattice is also the eigenstate of 
$\hat{\Delta}^2$ and $\hat{\Delta}_z$ with the eigenvalues $\Delta(\Delta+1)$ and $\Delta_z\,(=0$ at half filling), 
respectively. 
We can also find in Fig.~\ref{fig:structure_tr}(a) that the photoexcited state $|\Psi(t)\rangle$ acquires finite overlap 
$|\braket{\psi_m|\Psi(t)}|^2$ with the eigenstates $\ket{\psi_m}$ of $\hat{\cal{H}}$ with nonzero $\Delta$ 
[see also Fig.~\ref{fig:structure_tr}(b)]. 
These eigenstates $\ket{\psi_m}$ with nonzero $\Delta$ are photoexcited by transiently breaking the internal 
SU(2) structure during the pulse irradiation. 
This is exactly the reason for the enhancement of the electron-electron pair correlations in the photoexcited state 
$|\Psi(t)\rangle$. 

\subsection{Selection rule}\label{sec:sr}

The distribution of the weight $|\braket{\psi_m|\Psi(t)}|^2$ in the photoexcited state $\ket{\Psi(t)}$ 
among the eigenstates $\ket{\psi_m}$ found 
in Figs.~\ref{fig:structure}(a) and \ref{fig:structure_tr}(a) requires better understanding of  
the properties of the current operator $\hat{J}$ with respect to the $\Delta$-pairing operators. 
To be concrete, here we 
assume the 1D EFKM with $t^{(2)}_h = - t^{(1)}_h = t_h$ but the following argument is easily extended to 
other EFKMs, including 2D EFKM in the triangular lattice, as long as $t^{(2)}_h = - t^{(1)}_h = t_h$.  

In the 1D EFKM with the direct-gap-type band structure, i.e, $t^{(2)}_h = - t^{(1)}_h = t_h$, 
the current operator $\hat{J}^{(1)}_0 = \hat{J}$  is given as 
\begin{align}
\hat{J}^{(1)}_0   
\! =  \! i t_h \! \sum_{j} \! \sum_{\alpha=1,2} \! (-1)^{\alpha} \!  \left(\hat{c}^{\dag}_{j+1,\alpha} \hat{c}_{j,\alpha} - \hat{c}^{\dag}_{j,\alpha} \hat{c}_{j+1,\alpha}\right). 
\end{align}
We can now easily show that 
\begin{align}
\left[ \hat{\Delta}^\pm, \hat{J}^{(1)}_0 \right] = \sqrt{2} \hat{J}^{(1)}_{\pm 1} 
\label{eq:cr1}
\end{align}
and
\begin{align}
\left[ \hat{\Delta}_z, \hat{J}^{(1)}_0 \right] = 0, 
\label{eq:cr2}
\end{align} 
where $\hat{J}^{(1)}_{\pm 1}$ is defined as  
\begin{align}
\hat{J}^{(1)}_{+1} =\sqrt{2}i t_h \sum_j  \bigl( \hat{c}^\dag_{j,2} \hat{c}^\dag_{j+1,1} - \hat{c}^\dag_{j+1,2} \hat{c}^\dag_{j,1} \bigr) 
\end{align}
and
\begin{align}
\hat{J}^{(1)}_{-1} = \sqrt{2}i t_h \sum_j  \bigl(\hat{c}_{j+1,1} \hat{c}_{j,2}   - \hat{c}_{j,1} \hat{c}_{j+1,2} \bigr). 
\end{align}
We can also show that these two operators satisfy the following commutation relations: 
\begin{align}
\left[ \hat{\Delta}^\pm, \hat{J}^{(1)}_{\mp 1} \right] = \sqrt{2} \hat{J}^{(1)}_0 
\label{eq:cr3}
\end{align}
and 
\begin{align}
\left[ \hat{\Delta}_z, \hat{J}^{(1)}_{\pm 1} \right] = \pm \hat{J}^{(1)}_{\pm 1}. 
\label{eq:cr4}
\end{align} 
Note that to derive these commutation relations, 
we have not assumed any condition for the lattice system such as the bipartite lattice. 
This is in sharp contrast to the case of the $\eta$-pairing operators for the Hubbard model 
where the lattice must be bipartite to satisfy the similar commutation relations~\cite{KSSetal19}. 

From the commutation relations in Eqs.~(\ref{eq:cr1}), (\ref{eq:cr2}), (\ref{eq:cr3}), and (\ref{eq:cr4}), 
we can now immediately conclude that $\hat{J}^{(1)}_q$ 
with $q=0$, $\pm1$ is a rank-one tensor operator in terms of the $\Delta$-pairing operators.  
In particular, the current operator $\hat{J}^{(1)}_0 =\hat{J}$ 
is a rank-one tensor operator with $q=0$.
Therefore, according to the Wigner--Eckart theorem~\cite{JJSakurai,*MERose}, 
we have the following selection rule:  
\begin{align} 
\bra{\Delta',\Delta_z'}\hat{J}^{(1)}_0\ket{\Delta,\Delta_z} \propto
\left( \begin{array}{ccc}
\Delta & 1 & \Delta'           \\ 
\Delta_z & 0 &  -\Delta'_z  \\
\end{array} \right) 
\label{eq:3jrule}
\end{align} 
with the 3$j$-symbol, where $\ket{\Delta,\Delta_z}$ is the simultaneous eigenstate of 
$\hat{\Delta}$ and $\hat{\Delta}_z$. 
Since $\Delta_z'=\Delta_z=0$ at half filling, the selection rule becomes 
\begin{align}
\bra{\Delta', \Delta'_z \! = \! 0}\hat{J}^{(1)}_0 \ket{\Delta, \Delta_z \! = \! 0}  \ne  0
\end{align}
only for 
\begin{align}
\Delta'=\Delta\pm1 .
\label{eq:sr}
\end{align}

Based on this selection rule, the photoexcited processes in Figs.~\ref{fig:structure} and \ref{fig:structure_tr}
are understood as follows. 
In the small-$A_0$ limit, the external perturbation given in Eq.~(\ref{eq:phase}) 
is expressed as $A(t)\hat{J}$~\cite{LGBetal14},  
where $\hat{J}$ is the current operator defined above.   
Therefore, according to the selection rule in Eq.~(\ref{eq:sr}), 
in the linear response regime the photoinduced state $\ket{\Psi(t)}$ can contain 
the eigenstates $\ket{\psi_m}$ with $\Delta=1$ and the eigenenergies at $\varepsilon_m-\varepsilon_0 \sim U$, assuming that 
$\omega_p$ is tuned around $U$. 
This explains the good agreement between the optical spectrum $\chi_{JJ}(\omega)$ and $P(q\!=\! 0,t)$ 
found in Figs.~\ref{fig:optimal}(c) and \ref{fig:optimal_tr}(c). 
In the second order, the photoinduced state $\ket{\Psi(t)}$ can contain 
the eigenstates $\ket{\psi_m}$ with $\Delta=2$ at $\varepsilon_m-\varepsilon_0\sim 2U$, as well as 
$\Delta=0$ at $\varepsilon_m-\varepsilon_0\sim 0$ and $2U$. Applying the same argument for higher orders, 
the eigenstates $\ket{\psi_m}$ with even larger $\Delta$ values acquire in the transient period 
a finite overlap $|\braket{\psi_m|\Psi(t)}|^2$ with the photoinduced state $\ket{\Psi(t)}$. 
Considering all orders, 
eventually, the distribution of eigenstates $\ket{\psi_m}$ 
in the photoinduced state $\ket{\Psi(t)}$ forms a ``tower of states"~\cite{KSSetal19}, in which the eigenstates $\ket{\psi_m}$ 
with $\Delta$ even (odd) are excited at the excitation energy around 
$\varepsilon_m-\varepsilon_0\sim\,{\rm even\,(odd)\,\,integer}\times U$. 
In other words, the eigenstates $\ket{\psi_m}$ 
with $\Delta$ even (odd) are absent in the photoinduced state $\ket{\Psi(t)}$ at the excitation energy around 
$\varepsilon_m-\varepsilon_0\sim\,{\rm odd\,(even)\,\,integer}\times U$.
This is indeed in good qualitative accordance with the numerical results 
in Figs.~\ref{fig:structure}(a) and \ref{fig:structure_tr}(b).

\subsection{Different band width}

So far, we have assumed that $t_h^{(1)}=-t_h^{(2)}$. 
However, when the valence and conduction bands have different band widths, i.e. 
$t_h^{(1)} \ne -t_h^{(2)}$,  
the commutation relations with respect to the $\Delta$-pairing operators are broken because 
$[\hat{\mathcal{H}}_{0},\hat{\Delta}^{\pm}]  \ne 0$. 
Here, we investigate the electron-electron pair correlations in the photoexcited state 
when the internal SU(2) structure is broken in the EFKM. 

\begin{figure}[t]
\begin{center}
\includegraphics[width=0.9\columnwidth]{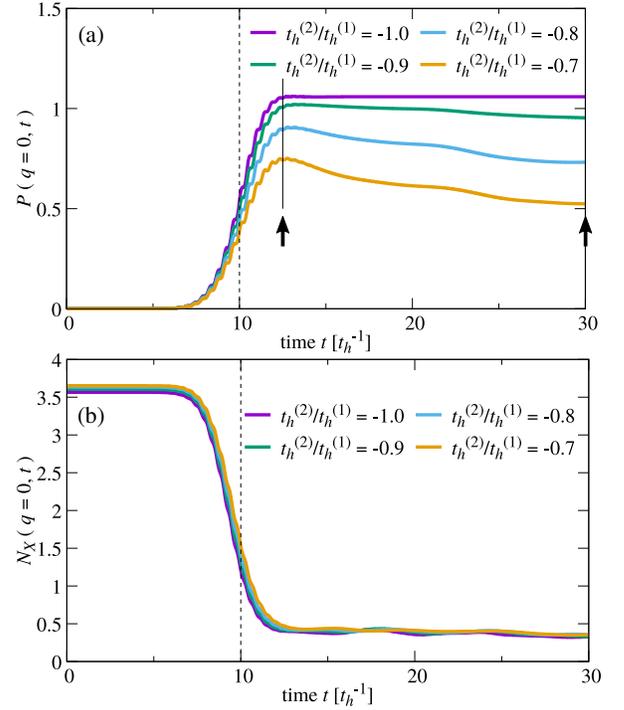}
\caption{
Time evolution of (a) the electron-electron pair structure factor $P(q = 0,t)$ and 
(b) the excitonic (i.e., electron-hole pair) structure factor $N_X(q = 0,t)$
for the 1D EFKM with different band widths ($t_h = |t_h^{(1)}| > |t_h^{(2)}|$) at half filling. 
The results are calculated by the ED method for $L=16$ at $U=8t_h$ under PBC.   
We set $D=0.75 t_h$, $0.65 t_h$, $0.6 t_h$, and $0.55 t_h$ for $t^{(2)}_h/t^{(1)}_h=-1.0$, $-0.9$, $-0.8$, 
and $-0.7$, respectively, in which $N_1=12$ and $N_2 =4$. 
The vector potential $A(t)$ is adopted with $A_0=0.4$, $\omega_p=7t_h$, $\sigma_p=2/t_h$, 
and $t_0=10/t_h$. The vertical dashed lines indicate $t_0$. 
}
\label{fig:dbw_corr}
\end{center}
\end{figure}

\begin{figure}[t]
\begin{center}
\includegraphics[width=\columnwidth]{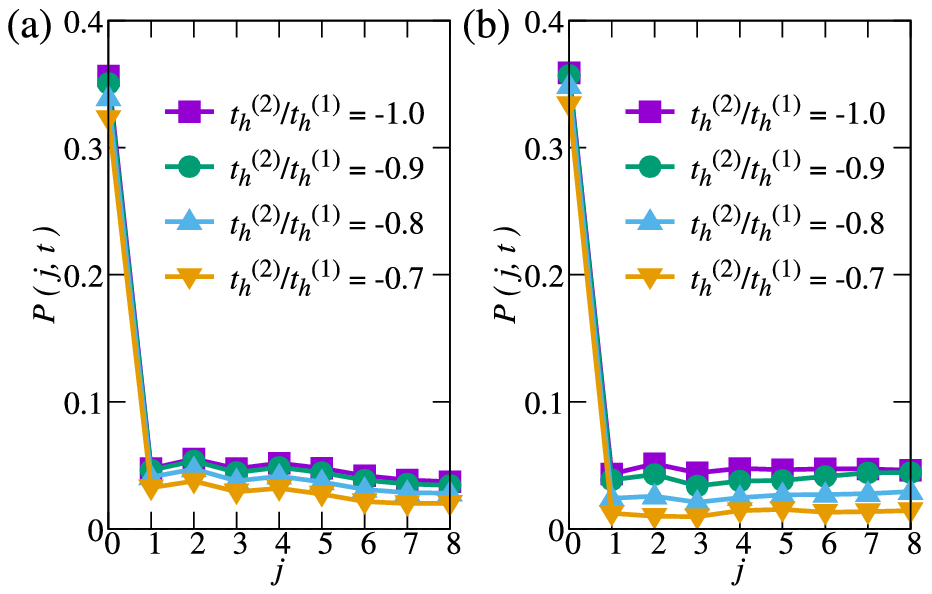}
\caption{
The on-site electron-electron pair correlation function $P(j,t)$ in the photoexcited state at 
(a) $t=12.5/t_h$ and (b) $t=30/t_h$ indicated by arrows in Fig.~\ref{fig:dbw_corr}(a). 
The results are for the 1D EFKM with $U=8t_h$ in $L=16$ at half filling. 
We set $D=0.75 t_h$, $0.65 t_h$, $0.6 t_h$, and $0.55 t_h$ 
for $t^{(2)}_h/t^{(1)}_h=-1.0$, $-0.9$, $-0.8$, and $-0.7$, respectively,  
in which $N_1=12$ and $N_2 =4$. 
The vector potential $A(t)$ is adopted with $A_0=0.4$, $\omega_p=7t_h$, $\sigma_p=2/t_h$, 
and $t_0=10/t_h$. 
}
\label{fig:dbw_corr_r}
\end{center}
\end{figure}

Figure~\ref{fig:dbw_corr} shows the time evolution of the electron-electron pair structure factor $P(q=0,t)$ 
and the excitonic structure factor $N_X(q=0,t)$ for the photoexcited state 
$|\Psi(t)\rangle$ with different values of $t_h^{(2)}/ t_h^{(1)}$ in the 1D EFKM. 
Although the internal SU(2) structure with respect to the $\Delta$-pairing operators is broken when $t_h^{(1)} \ne -t_h^{(2)}$, 
we find the enhancement of the electron-electron pair correlations (see also Fig.~\ref{fig:dbw_corr_r}). 
Note that $P(q = 0,t)$ is no longer conserved after the pulse irradiation when $t_h^{(1)} \ne -t_h^{(2)}$ 
because of $[\hat{\mathcal{H}}_{0},\hat{\Delta}^{+}\hat{\Delta}^{-}] \ne 0$.
With decreasing $|t_h^{(2)}/ t_h^{(1)}|$, $P(q=0 ,t)$ is more suppressed after the pulse irradiation. 
However, as shown in Fig.~\ref{fig:dbw_corr_r}, the on-site electron-electron pair correlations in the photoexcited state 
are still robust specially in the transient period. 

\section{Conclusion}\label{sec:summary}

We have investigated the photoinduced electron-electron pairing in the half-filled EFKM with the 
direct-gap-type band structure. 
By employing the time-dependent ED method, we have shown the enhancement of the on-site electron-electron 
pair correlations with the corresponding pair structure factor exhibiting a sharp peak at $\bm{q}=\bm{0}$ in the 
photoexcited state, while the initial ground state excitonic (i.e., electron-hole pair) correlations are strongly suppressed. 
We have shown that there exists the internal SU(2) structure with respect to the $\Delta$-pairing operators in 
the EFKM $\hat{\cal{H}}$ with the direct-gap-type band structure, i.e., $t_h^{(1)}=-t_h^{(2)}$, and therefore 
any eigenstate of $\hat{\cal{H}}$ can be simultaneously the eigenstate of the $\Delta$-pairing operators, 
characterizing the number of $\Delta$ pairs.  
The analysis for the distribution of the eigenstates of $\hat{\cal{H}}$ in the photoexcited state reveals that 
the photoexcited state captures nonzero weight of the eigenstates of $\hat{\cal{H}}$ that possess a finite number of 
$\Delta$ pairs. This is the essential reason for the enhancement of the on-site electron-electron pair correlations 
in the photoexcited state.

The internal SU(2) relations with respect to the $\Delta$-pairing operators are preserved even for the EFKM on 
nonbipartite lattices 
such as the triangular lattice, in which the on-site electron-electron pairing with momentum $\bm{q}=\bm{0}$ can also be 
photoinduced in the EFKM with the direct-type band structure, i.e., $t_h^{(1)}=-t_h^{(2)}$. 
This is in sharp contrast to the photoinduced $\eta$-pairing in the repulsive Hubbard model, for which the bipartite 
lattices are required to preserve the internal SU(2) structure with respect to the $\eta$-pairing operators. 
We have also shown that the photoinduced states still displays the robust on-site electron-electron pairing correlations 
even when the internal SU(2) structure is broken by setting the different band widths of the valence and conduction 
bands, i.e., $t_h^{(1)}\ne-t_h^{(2)}$, as long as $|t_h^{(1)}/t_h^{(2)}|$ is close to one.
Although we have shown the enhancement of the electron-electron pair correlation in the 
relatively small finite size clusters that can be treated by the ED method, we expect the similar enhancement 
even in larger clusters. This is simply because the previous matrix-product state calculations for the 1D Hubbard 
model have clearly found the photoinduced enhancement of the $\eta$-pairing correlation 
in larger clusters~\cite{KSSetal19}.
However, we should also note that in order for the photoinduced state to exhibit the long-range superconducting order, 
i.e., the electron-electron pair structure factor $P(q=0)/L$ being finite in the thermodynamic limit, 
the $\Delta$-pairing state with $\Delta$ proportional to the system size $L$ has to be photoexcited 
[e.g., see Eq.~(\ref{eq:Pq=0})]. 

The recent experimental observation of photoinduced superconductivity and increase of superconducting 
transition temperature in some of 
high-$T_{\rm c}$ cuprates~\cite{FTDetal11,HKNetal14,KHNetal14} and alkali-doped 
fullerenes~\cite{MCNetal16,CBJetal18} has stimulated extensive theoretical studies of light induced 
superconductivity~\cite{SKGetal16,KBRetal16,KWRetal17,STGetal17,Se17,MTEetal17,BTKetal19}. 
The main focus in these theoretical studies is a photoinduced state with physical properties that is already 
present in the corresponding equilibrium phases. 
In contrast, the enhancement of electron-electron pair correlations found in our study cannot be simply explained by a 
dynamical transition that is induced by effectively varying the model parameters because there is no region 
in the ground state phase diagram of the EFKM showing large electron-electron pairing correlations even away from 
half filling. 
Therefore, our finding is distinct from the previous theoretical studies and provides a new insight into 
photoinduced phenomena. 


In this paper, we have focused on the time-dependent correlation functions. 
However, the time-dependent dynamical spectra such as the time-resolved optical 
conductivity~\cite{FCNetal12,LGBetal14}, angle-resolved photoemission 
spectroscopy~\cite{FKP09,SKMetal13,SCKetal15}, and resonant inelastic x-ray scattering~\cite{CWJetal19} 
might provide deeper understanding of a photoinduced state. 
Moreover, the EFKM considered in this paper is the {\it spinless} model. 
The realistic models for possible excitonic materials should have the spin degrees of freedom, 
and thus our theory has to be extended to a {\it spinful} model such as the two-band Hubbard 
model~\cite{KSO12,KA14-1,FSO18,NMKetal19}. 
Furthermore, the importance of the electron-phonon coupling has been pointed out in the excitonic candidate 
materials TiSe$_2$ and Ta$_2$NiSe$_5$~\cite{WRCetal11,KZFetal15,NHTetal18}. 
Therefore, in order to understand the pump-probe experiments reported recently in these materials, 
the phonon degrees of freedom are also important in the theory. 
These are intriguing extensions of the present study in the future.


\begin{acknowledgments}
The authors acknowledge T. Shirakawa and K. Sugimoto for fruitful discussion. 
This work was supported in part by Grants-in-Aid for Scientific Research from JSPS 
(Projects No.~JP17K05530, No.~JP18H01183, No.~JP18K13509, and No.~JP19J20768) of Japan.
\end{acknowledgments}

R.F. and T.K. contributed equally to this work.

\begin{appendix}

\section{Photoinduced $\eta$-pairing in EFKM} \label{sec:app-eta}

In this appendix, we discuss the electron-electron pairing in the EFKM $\hat{\cal{H}}$ with the indirect-gap-type 
band structure~\cite{direct}. 
First, we introduce the interorbital $\hat{\eta}$-pairing operators defined as 
\begin{align}
\hat{\eta}^+_j=  (-1)^j \hat{c}^{\dag}_{j,2} \hat{c}^{\dag}_{j,1}, ~~\hat{\eta}^-_j=  (-1)^j \hat{c}_{j,1} \hat{c}_{j,2}
\end{align}
and 
\begin{align}
\hat{\eta}^z_j=\frac{1}{2}  \left( {\hat n}_{j, 1} + {\hat n}_{j, 2} -1 \right), 
\end{align}
which satisfy the SU(2) commutation relations, i.e., 
\begin{align}
&\left[ \hat{\eta}^+_j, \hat{\eta}^-_j \right] = 2 \hat{\eta}^z_j , 
\\
&\left[ \hat{\eta}^z_j,  \hat{\eta}^\pm_j \right] = \pm \hat{\eta}^\pm_j .
\end{align}
The total $\hat{\eta}$ operators, $\hat{\eta}^\pm=\sum_j\hat{\eta}^\pm_j$ and $\hat{\eta}_z=\sum_j\hat{\eta}^z_j$,  
also satisfy the SU(2) commutation relations.  

\begin{figure}[!t]
\begin{center}
\includegraphics[width=0.99\columnwidth]{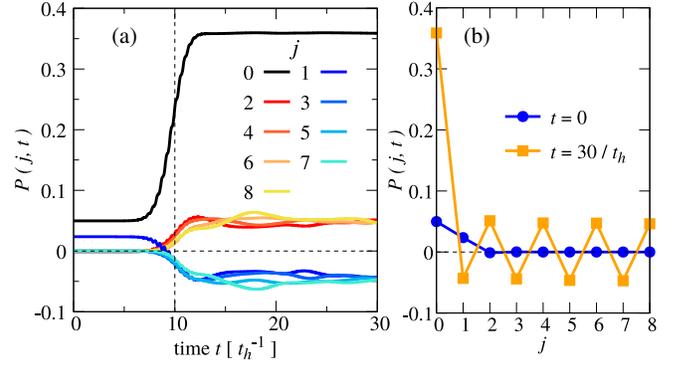}
\caption{
(a) Time evolution of the on-site electron-electron pair correlation function $P(j,t)$ and  
(b) $P(j,t)$ at $t=0$ (blue circles) and $t=30/t_h$ (orange squares).  
The results are for the 1D EFKM under PBC with $t_h^{(1)}=t_h^{(2)}=t_h$, $U=8t_h$, and $D=0.75 t_h$ in 
$L=16$, for which $N_1=12$ and $N_2 =4$. 
We set $A_0=0.4$, $\omega_p=7t_h$, $\sigma_p=2/t_h$, and $t_0=10/t_h$ for the vector potential $A(t)$ 
defined in Eq.~(\ref{A(t)}).  
}\label{fig:pair_corr_eta}
\end{center}
\end{figure}

The important property of the $\eta$-pairing operators is 
\begin{align}
\left[ \hat{\mathcal{H}}_0,\hat{\eta}^{+} \right]  
=  \sum_{\bm{k}} \left[ \epsilon_{1}(\bm{k})   + \epsilon_{2}(\bm{\pi}-\bm{k})  \right] \hat{c}^{\dag}_{\bm{\pi}-\bm{k},2} \hat{c}^{\dag}_{\bm{k},1}, 
\label{PI-EFKM_eq2}   
\end{align}
where $\hat{\mathcal{H}}_0=\sum_{\bm{k},\alpha}\epsilon_{\alpha}(\bm{k})c^{\dag}_{\bm{k},\alpha}c_{\bm{k},\alpha}$ 
and $\bm{\pi} = (\pi,\cdots,\pi)$.  
For the $d$-dimensional cubic lattice, for example, 
$\epsilon_{2}(\bm{\pi}-\bm{k}) = - \epsilon_{2}(\bm{k})$ and therefore  
the commutation relation becomes 
\begin{align}
\left[ \hat{\mathcal{H}}_0,\hat{\eta}^{+} \right]=-2(t^{(1)}_h-t^{(2)}_h)
\sum_{\tau, \bm{k}}\cos(k_{\tau}) \, \hat{c}^{\dag}_{\bm{\pi}-\bm{k},2} \hat{c}^{\dag}_{\bm{k},1}.
\end{align}
Note that this commutation relation cannot be satisfied in the triangular lattice 
because $\epsilon_{2}(\bm{\pi}-\bm{k}) \ne - \epsilon_{2}(\bm{k})$. This is in sharp contrast to the case of 
the $\Delta$-pairing operators, for which the corresponding commutation relation in Eq.~(\ref{PI-EFKM_eq_H-Delta}) 
is satisfied even for the EFKM in nonbipartite lattices such as the triangular lattice.  
A similar relation is also satisfied for $\hat{\eta}^{-}$.  
Thus, in the $d$-dimensional bipartite cubic lattice, we have the relation $[ \hat{\mathcal{H}}_0,\hat{\eta}^{\pm} ]=0$ 
when $t^{(1)}_h = t^{(2)}_h$.  
We can also show that $[\hat{\mathcal{H}}_U,\hat{\eta}^{\pm}] = \pm U \hat{\eta}^{\pm}$. 
Therefore, we obtain the following relation:  
\begin{align}
\left[ \hat{\mathcal{H}},\hat{\eta}^{\pm} \right] = \pm U \hat{\eta}^{\pm}
\end{align}
for the EFKM when $\epsilon_{2}(\bm{\pi}-\bm{k}) = - \epsilon_{1}(\bm{k})$. 
It is easily shown that the same commutation relations are satisfied more generally for the EFKM in any 
bipartite lattice, including the honeycomb lattice, as long as $t^{(1)}_h = t^{(2)}_h$.
Notice that these relations are essentially the same as those found in the Hubbard model~\cite{Ya89,EFGetal05}. 
This is understood simply because the EFKM is exactly the same as the Hubbard model with the Zeeman term 
when $t^{(1)}_h = t^{(2)}_h$.  

Consequently, introducing  
\begin{align}
{\hat{\eta}}^2=\frac{1}{2}\left( \hat{\eta}^+\hat{\eta}^- + \hat{\eta}^-\hat{\eta}^+  \right) + \hat{\eta}_z^2,
\end{align} 
we have 
\begin{align}
\left[ \hat{\cal{H}}, \hat{\eta}^2 \right] = \left[ \hat{\cal{H}}, \hat{\eta}_z \right]=0
\end{align}
for the EFKM with $ t^{(1)}_h = t^{(2)}_h$ in the bipartite lattice. 
Thus, any eigenstate of $\hat{\cal{H}}$ is also the eigenstate of ${\hat{\eta}}^2$ and $\hat{\eta}_z$ 
with the eigenvalues $\eta(\eta+1)$ and $\eta_z$, respectively.  
We therefore expect that the density-wave-like pair correlations 
are enhanced by the pulse irradiation~\cite{KSSetal19}.  

Figure~\ref{fig:pair_corr_eta}(a) shows the time evolution of the real-space electron-electron pair correlation function 
$P(j,t)$ in the 1D EFKM with $t_h^{(1)}=t_h^{(2)}=t_h$.
$P(j,t)$ at $j=0$ corresponding to the double occupancy $n_d (t)$ is enhanced by pulse irradiation. 
$P(j \! \ne \! 0,t)$ is also enhanced significantly by the pulse irradiation, similar to Fig.~\ref{fig:pair_corr}(a), 
but now oscillates with the opposite phases between odd and even sites. 
As shown in Fig.~\ref{fig:pair_corr_eta}(b), the pair correlation after the pulse irradiation extends to longer distances 
over the cluster, as compared to that of the initial state before the pulse irradiation. 
It is also clear that the sign of $P(j,t)$ alternates between neighboring sites 
and we confirm that $P(j,t)$ in Fig.~\ref{fig:pair_corr_eta}(b) is consistent with $(-1)^{j}P(j,t)$ 
in Fig.~\ref{fig:pair_corr}(b).
Therefore, in the indirect-gap-type band system, the $\eta$-pairing correlation is 
enhanced by the pulse irradiation.

\end{appendix}


\bibliography{References}

\begin{thebibliography}{87}
\bibitem [{dir()}]{direct}%
  \BibitemOpen
  \href@noop {} {}\bibinfo {note} {Direct (indirect) implies that the valence
  band maximum and the conduction band minimum are located at the same
  (different) position(s) in the Brillouin zone.}\BibitemShut {Stop}%
\bibitem [{SU2()}]{SU2}%
  \BibitemOpen
  \href@noop {} {}\bibinfo {note} {In order for $\hat{\mathcal{H}}$ to be fully
  rotation symmetric with respect to the $\Delta$-pairing operators,
  $\hat{\mathcal{H}}$ has to satisfy $[ \hat{\mathcal{H}},\hat{\Delta}^{\pm} ]
  =0$, instead of Eq.~(\ref{PI-EFKM_eq_HEFKM-Delta}), in addition to $[
  \hat{\mathcal{H}}, \hat{\Delta}_z ]=0$. If we add the on-site energy term
  $-\frac{U}{2}\sum_j ( \hat{n}_{j,1} + \hat{n}_{j,2} )$ to the EFKM in
  Eq.~(\ref{PI-EFKM_eq_H}), the model satisfies these commutation relations and
  thus preserves the internal SU(2) symmetry with respect to the
  $\Delta$-pairing operators. Note however that the additional term,
  $-\frac{U}{2}\sum_j ( \hat{n}_{j,1} + \hat{n}_{j,2} )$, is not important for
  our purpose because the eigenstate of $\hat{\mathcal{H}}$ is still
  simultaneously the eigenstate of $\hat{\Delta}^2$ and $\hat{\Delta}_z$,
  irrespectively of the additional on-site energy term.}\BibitemShut {Stop}%
\bibitem [{del()}]{delta}%
  \BibitemOpen
  \href@noop {} {}\bibinfo {note} {$\hat{\Delta}$ and $\hat{\Delta}_z$ with the
  hat are operators and $\Delta$ and $\Delta_z$ without the hat are
  $c$-numbers.}\BibitemShut {Stop}%
\end{thebibliography}

\end{document}